\documentclass[a4paper,12pt]{article}
\usepackage[a4paper,text={16.8cm,22.4cm}]{geometry}
\usepackage{amsmath,amssymb,latexsym,bm,braket}

\usepackage{graphicx}

\usepackage{cite}

\allowdisplaybreaks 
\usepackage{color}

\newcommand{\Tr}{\mathrm{Tr}}

\newcommand{\be}{\begin{equation}}
\newcommand{\ee}{\end{equation}}
\newcommand{\bea}{\begin{eqnarray}}
\newcommand{\eea}{\end{eqnarray}}
\newcommand{\sqrts}{\sqrt{\hat{s}}}
\newcommand{\cusp}{\mbox{{\tiny cusp}}}

\newcommand{\msbar}{\overline{\text{MS}}}

\begin{document}

\begin{titlepage}

  \begin{flushright}
DCPT/13/164\\
  IPPP/13/82 
  \end{flushright}
  
  \vspace{5ex}
  
  \begin{center}
    \boldmath
    \textbf{\Large Boosted top production: factorization and resummation for single-particle inclusive distributions} \vspace{7ex}
    \unboldmath    
    
    \textsc{Andrea Ferroglia$^b$, Simone Marzani$^c$, Ben D. Pecjak$^c$, and Li Lin
      Yang$^{a,d,e}$}

    \vspace{2ex}
  
    \textsl{${}^a$School of Physics and State Key Laboratory of Nuclear Physics and Technology,\\
      Peking University, Beijing 100871, China
      \\[0.3cm]
      ${}^b$New York City College of Technology, 300 Jay Street\\
      Brooklyn, NY 11201, USA
      \\[0.3cm]
      ${}^c$Institute for Particle Physics Phenomenology,
      University of Durham\\
      DH1 3LE Durham, UK
      \\[0.3cm]
      ${}^d$Collaborative Innovation Center of Quantum Matter, Beijing, China
      \\[0.3cm]
      ${}^e$Center for High Energy Physics, Peking University, Beijing 100871, China}
  \end{center}

  \vspace{4ex}

  \begin{abstract}
    
    We study single-particle inclusive (1PI) distributions in
    top-quark pair production at hadron colliders, working in the
    highly boosted regime where the top-quark $p_T$ is much larger
    than its mass.  In particular, we derive a novel factorization
    formula valid in the small-mass and soft limits of the
    differential partonic cross section. This provides a framework for
    the simultaneous resummation of soft gluon corrections and
    small-mass logarithms, and also an efficient means of obtaining
    higher-order corrections to the differential cross section in 
    this limit. The result involves five distinct one-scale functions,
    three of which arise through the subfactorization of soft real
    radiation in the small-mass limit. We list the NNLO corrections to
    each of these functions, building on results in the literature by
    performing a new calculation of a soft function involving four
    light-like Wilson lines to this order.  We thus obtain a nearly
    complete description of the small-mass limit of the differential
    partonic cross section at NNLO near threshold, missing only terms
    involving closed top-quark loops in the virtual corrections.

\end{abstract}

\end{titlepage}

\section{Introduction}
\label{sec:intro}

Nowadays, top-quark production is of great interest in elementary
particle phenomenology at hadron colliders. This is due to the fact
that top-quark physics is closely connected to the study of the
recently discovered Higgs boson \cite{Aad:2012tfa, Chatrchyan:2012ufa}
and to the search for new particles. Millions of top-quark pair events
have already been produced at the Large Hadron Collider (LHC). For
this reason, the ATLAS and CMS collaborations were able to measure the
top-quark pair production cross section with remarkable precision,
e.g.~\cite{Aad:2012mza, Aad:2012vna, ATLAS:2012al, Khachatryan:2010ez,
  Chatrchyan:2013ual, Chatrchyan:2013kff}.  On the theoretical side,
precise measurements require calculations of the measured observables
which include corrections beyond the next-to-leading-order (NLO) in
QCD. As an example, the total cross section, which can be measured
with a relative error of approximately $5 \%$, was recently evaluated
at next-to-next-to-leading order (NNLO) in perturbation theory
\cite{Czakon:2013goa}.

Differential distributions, such as the pair invariant mass
distribution, the top-quark rapidity distribution, and the
distributions with respect to the transverse momentum (both of the
individual top quark or of the $t \bar{t}$ system) are also of great
interest, especially in the search for new physics. The ATLAS and CMS
collaborations already measured several differential distributions
\cite{Aad:2012hg, Chatrchyan:2012saa}.  To date, the full set of NNLO
QCD corrections to these observables is not known. However, studies of
the soft gluon emission corrections to the top-pair invariant mass
distribution up to next-to-next-to-leading-logarithmic (NNLL) accuracy
were presented in \cite{Ahrens:2009uz, Ahrens:2010zv}. In those works,
the resummation of the soft corrections was carried out in momentum
space by employing methods developed in \cite{Becher:2006nr,
  Becher:2006mr, Becher:2007ty}. In the same papers, approximate
formulas including all of the terms proportional to logarithmic (plus
distribution) corrections up to NNLO were obtained starting from the
NNLL resummation formulas. A study of the top-quark transverse
momentum and rapidity distributions within the same approach was
carried out in \cite{Ahrens:2011mw}\footnote{Approximate NNLO formulas
  for the same observables obtained by means of standard Mellin space
  resummation methods can be found in \cite{Kidonakis:2010dk,
    Kidonakis:2011zn}.}.  The NNLL resummation of the transverse
momentum distribution of the $t \bar{t}$ system, which presents
additional technical complications with respect to the two
distributions mentioned above, was considered in \cite{Zhu:2012ts,
  Li:2013mia}.

A kinematic situation of special interest for new physics searches is
the so-called boosted regime, where the top quarks are produced with
energies much larger than their mass. Examples of boosted top
production include the differential distribution at high values of
pair invariant mass $M$, or the high-$p_T$ tail of the top-quark
transverse momentum distribution. The presence of new heavy particles
decaying into pairs of energetic top quarks could generate bumps or
more subtle distortions of differential distributions in this
kinematic region. The LHC at center-of-mass energies of $7$~TeV and
$8$~TeV has started to explore boosted top production experimentally,
and more data will become available with the future 14~TeV run. At the
same time, highly-boosted production is characterized by energy scales
much larger than the top-quark mass, and QCD calculations of the
differential cross sections must take this into account.

A factorization formalism appropriate for describing QCD corrections
to the pair invariant mass distribution in the limit $M\gg m_t$ was
put forth in \cite{Ferroglia:2012ku}, opening up the opportunity to
resum simultaneously soft-gluon corrections and small-mass logarithms
of the form $\ln(m_t/M)$. This same formalism can be used as a way of simplifying
the calculation of higher-order corrections in the small-mass limit, a
fact exploited in \cite{Ferroglia:2013zwa} to obtain an NNLO soft plus
virtual approximation to the invariant mass distribution in this
limit. The goal of the present work is to develop the framework
necessary for describing the highly-boosted limit of single-particle
inclusive (1PI) distributions, for instance the $p_T \gg m_t$ region
of the top-quark transverse momentum distribution. To this end, we
derive a factorization formalism appropriate for describing the double
soft and small-mass limit of the differential partonic cross section.

Our results have some common ground with those for the pair
invariant mass distribution in \cite{Ferroglia:2012ku}, a fact which
is particularly clear when deriving results in the double soft and
small-mass limit using those in the soft limit as a starting point.
In the soft limit, the partonic cross section for top-quark pair
production factorizes into a hard function, related to virtual
corrections, and a soft function, related to real emission in the soft
limit \cite{Kidonakis:1996aq, Kidonakis:1997gm,
  Banfi:2005cv, Kidonakis:2000ui, Kidonakis:2001nj, Kidonakis:2003qe,
  Kidonakis:2008mu, Kidonakis:2011zn, Kidonakis:2010dk, Ahrens:2010zv,
  Ahrens:2011mw}.  The hard function is common
to both cases, while the soft function depends on the observable.
The small-mass factorization of the virtual corrections to 1PI
observables can thus be taken directly from \cite{Ferroglia:2012ku}.
It involves the virtual corrections calculated with
$m_t=0$, in the form of a ``massless hard function'', and a 
second function encoding all $m_t$ dependence and related to collinear
divergences in the small-mass limit.

On the other hand, the small-mass factorization of soft real radiation
for 1PI observables in top-quark pair production has not yet been 
discussed in the literature\footnote{Note, however, that the 
single-hadron inclusive cross section at large values of the 
transverse momentum of the produced hadron 
was recently studied in \cite{Catani:2013vaa}.}, and
turns out to be rather different than that for the pair invariant mass
distribution.  A main result of our paper is that such real radiation
factorizes into three component functions, as shown in
(\ref{eq:softfact}) below.  The physical interpretation is that soft
radiation collinear to the observed top quark, soft radiation
collinear to the unobserved anti-top quark, and wide angle soft
emission are decoherent and factorize.  We make a technical
distinction between these different kinds of soft radiation in two
ways: diagrammatically, through the method of regions, and at the
operator level, in terms of Wilson loops. Our final results associate
{\it i}) wide-angle soft emission with a Wilson loop built out of four
light-like Wilson lines and involving a delta-function constraint
particular to 1PI observables; {\it ii}) soft radiation collinear to
the top quark with the Wilson loop defining the soft part of the
heavy-quark fragmentation function \cite{Mele:1990cw} (this is
equivalent to the partonic shape-function from $B$ meson decays); and
{\it iii}) soft radiation collinear with the anti-top quark with the
Wilson loop defining the heavy-quark jet function introduced in
\cite{Fleming:2007qr}.  While the ``massless'' soft function involving
four light-like Wilson lines is a matrix in color space, the two types
of soft-collinear objects are color diagonal.

With this factorization at hand, one can resum soft and small-mass
logarithms at the level of the differential partonic cross section by
deriving and solving renormalization-group (RG) equations for the five
component functions, or else use it as a tool for simplifying the
calculation of higher-order corrections in this limit. In fact, of the
five component functions mentioned above, only the massless soft
function has not yet been calculated to NNLO; we build on the
literature by performing this computation here. We thus achieve a
nearly complete NNLO ``soft plus virtual'' approximation to the
differential partonic cross section in the small-mass limit.  The
final missing piece is the NNLO virtual corrections involving closed
heavy-quark loops and proportional to powers of $n_h=1$ for the top
quark.  We leave an analysis of these corrections to future work,
emphasizing their potential complications on the factorization
formalism in the small-mass limit.  Even in their absence, our results
represent the most complete fixed-order calculation of the $p_T$
distribution for boosted production performed so far. They go beyond
the approximate NNLO formulas derived in \cite{Ahrens:2011mw,
  Kidonakis:2010dk} by determining the non-logarithmic
(delta-function) coefficient in addition to the logarithmic plus
distribution terms.  They are also consistent with them, and the fact
that the NNLO logarithmic plus distribution contributions obtained with the
two methods are identical in the small-mass limit is a strong check on
our factorization formalism.

The paper is organized as follows. In Section~\ref{sec:kinandfac} we
introduce some notation and provide the factorization formula for the
partonic cross section in the double small-mass and soft limit.  We
then devote Section~\ref{sec:factorization} to details of factorizing
soft real radiation in the small-mass limit. In Section~\ref{sec:phsp}
we analyze NLO phase space integrals in this limit using the method of
regions, identifying three distinct momentum configurations which
appear at leading power. In Section~\ref{sec:facmasssoft} we discuss
the all-order factorization of the soft function into a convolution of
component parts related to these three momentum regions.  In
Section~\ref{sec:resum} we give expressions needed for fixed-order
expansions and present RG equations needed to resum logarithmic
corrections.  We discuss subtleties related to closed top-quark loops
in Section~\ref{sec:top_loops}, and conclude in
Section~\ref{sec:conclusions}.  Some details of the NNLO calculation
of the massless soft function for 1PI kinematics are given in
Appendix~\ref{sec:SoftFunction}, while explicit expressions for the
anomalous dimensions and matching coefficients are collected in
Appendix~\ref{sec:coeffs}.

\section{Kinematics and factorization \label{sec:kinandfac}}
We consider the scattering process
\begin{align}
  \label{eq:process}
  N_1(P_1) + N_2(P_2) \to t(p_3) + \bar{t}(p_4) + X \, ,
\end{align}
where $N_1$ and $N_2$ indicate the incoming protons (at the LHC) or
proton and anti-proton (at the Tevatron), while $X$ represents an
inclusive hadronic final state. In the Born approximation and also to
leading order in the soft limit we will deal with later on, two
different production channels contribute to the partonic scattering
process (\ref{eq:process}): the quark-antiquark annihilation and
gluon fusion channels.  The partonic processes which we will analyze
in detail are thus
\begin{align}
  \label{eq:partprocess}
  q(p_1) + \bar{q}(p_2) &\to t(p_3) + \bar{t}(p_4) + \hat{X}(k) \, , \nonumber
  \\
  g(p_1) + g(p_2) &\to t(p_3) + \bar{t}(p_4) + \hat{X}(k) \, ,
\end{align}
where $\hat{X}$ contains any number of emitted partons. The relations
between the hadronic momenta ($P_i$) and the momenta of the incoming
partons ($p_i$) are $p_1=x_1P_1$ and $p_2=x_2P_2$. At the hadronic
level, we define the Mandelstam variables as
\begin{align}
  \label{eq:mandelstam}
  s &= (P_1+P_2)^2 \, , \quad t_1 = (P_1-p_3)^2-m_t^2 \, , \quad u_1=(P_2-p_3)^2 -m_t^2 \,
  ,
\end{align}
while the corresponding quantities at the partonic level are given by
\begin{gather}
  \hat{s} = x_1 x_2 s \, , \quad \hat{t}_1 = x_1 t_1 \, , \quad \hat{u}_1= x_2 u_1 \, .
   \label{eq:mandelstampart}
  \end{gather}
 It will also be useful to introduce the variable
  \begin{align}
   \quad s_4 = \hat{s} + \hat{t}_1 + \hat{u}_1 = (p_4+k)^2 - m_t^2 \,.
  \, 
\end{align}
Momentum conservation implies that $s_4 = 0$  at Born level ($k=0$).   
  
We will be interested in the double differential distribution with
respect to the transverse momentum $p_T$ and rapidity $y$ of the top
quark in the laboratory frame.  Such 1PI observables are obtained by
integrating over the phase space of the unobserved anti-top quark, along
with any extra real radiation.  The $p_T$ and rapidity are related to
the hadronic invariants (\ref{eq:mandelstam}) according to
\begin{align}
  t_1 = -\sqrt{s} \, m_\perp \, e^{-y} \, , \quad u_1 = -\sqrt{s} \, m_\perp e^{y} \, ,
\end{align}
where $m_\perp=\sqrt{p_T^2+m_t^2}$. Using (\ref{eq:mandelstampart})  allows one to express
the partonic Mandelstam variables in terms of the $p_T,y,x_1,x_2$.  Then, assuming factorization
in QCD\footnote{We should mention however potential subtleties in two-to-two processes pointed out in \cite{Forshaw:2006fk, Forshaw:2008cq, Catani:2011st, Forshaw:2012bi}.} and ignoring power corrections in $\Lambda_{\rm QCD}/m_t$, one can write the double differential distribution as 
\begin{align}
  \label{eq:pt,y}
  \frac{d\sigma}{dp_Tdy} &= \frac{16\pi p_T}{3s} \sum_{i,j} \int_{x_1^{\text{min}}}^1
  \frac{dx_1}{x_1} \int_{x_2^{\text{min}}}^{1} \frac{dx_2}{x_2} \, f_{i/N_1}(x_1,\mu_f) \,
  f_{j/N_2}(x_2,\mu_f) \, C_{ij}(s_4,\hat{s},\hat{t}_1,\hat{u}_1,m_t,\mu_f) \, ,
\end{align}
where the $f_{i/N}$ are universal non-perturbative PDFs for the parton
$i$ in the hadron $N$ and the hard-scattering kernels $C_{ij}$ are
related to the partonic cross section and can be calculated
perturbatively as series in the strong coupling constant.  In addition,
the lower limits of integration are given by 
\begin{align}
  x_1^{\text{min}} = \frac{-u_1}{s+t_1} \, , \quad x_2^{\text{min}} =
  \frac{-x_1t_1}{x_1s+u_1} \, .
\end{align}

The hard-scattering kernel is a function of the kinematic invariants
needed to describe the differential cross section.  As long as these
invariants are parametrically of the same order, an expansion of the
$C_{ij}$ in fixed orders of the strong coupling constant is
appropriate.  An interesting situation arises when there is a large
hierarchy among two or more of the kinematic invariants.  In that case
it is often possible to factorize the hard-scattering kernel into a
product of simpler functions depending only on a single mass scale, up
to corrections in the small ratio of disparate scales.  This
factorization is useful for two reasons.  First, the component
functions are typically easier to calculate than the full hard-scattering 
kernels.  Second, the factorization formula can be used as a
starting point for resumming large logarithmic corrections in the
ratio of scales which appear in the higher-order perturbative corrections.

An example often considered in the literature is the soft gluon
emission limit, where the partonic invariants satisfy the parametric
relation $s_4 \ll m_t^2, \hat{s},\hat{t}_1,\hat{u}_1$.  In this limit
the $C_{ij}$ factorize into a matrix product of a hard function
$\bm{H}^m_{ij}$ and a soft function $\bm{S}^m_{ij}$ as
follows:\footnote{The superscript $m$ in (\ref{eq:fac}) indicates that
  the hard and soft function are evaluated as exact functions of
  $m_t$, as opposed to the corresponding functions calculated with
  $m_t=0$ and used below.}
\begin{align}
  \label{eq:fac}
  C_{ij}(s_4,\hat{s},\hat{t}_1,\hat{u}_1,m_t,\mu_f) = \Tr \left[
    \bm{H}^m_{ij}(\hat{s},\hat{t}_1,\hat{u}_1,m_t,\mu_f) \,
    \bm{S}^m_{ij}(s_4,\hat{s},\hat{t}_1,\hat{u}_1,m_t,\mu_f) \right] + \mathcal{O}\left(\frac{s_4}{m_t^2}\right)
  \, .
\end{align}
Such a factorization was first derived in \cite{Laenen:1998qw}.  The
hard and soft functions are two-by-two matrices for the $q\bar q$
channel, and three-by-three for the $gg$ channel; the matrix structure
is related to the mixing of a basis of color-singlet amplitudes
through soft gluon exchange.  The hard function is related to virtual
corrections, and the soft function is related to real emission in the
soft limit. Real emission in the soft limit is considerably easier to
calculate than in the generic case. The eikonal factors related to
soft gluon emissions exponentiate into Wilson lines, and at the level
of the squared matrix element form a gauge invariant Wilson loop
operator.  Much is known about the perturbative properties of such
Wilson loops. In position, Laplace, or Mellin space they contain a
series of double logarithmic corrections.  In momentum space, these
translate into logarithmic plus distribution and delta-function
corrections.  In particular, defining expansion coefficients of the
hard-scattering kernels as
\begin{align}
C_{ij}(s_4,\hat{s},\hat{t}_1,\hat{u}_1,m_t,\mu_f) = \alpha_s^2\bigg[ &
C^{(0)}_{ij}(s_4,\hat{s},\hat{t}_1, \hat{u}_1,m_t,\mu_f)+ \frac{\alpha_s}{4\pi} C^{(1)}_{ij}(s_4,\hat{s},\hat{t}_1,\hat{u}_1,m_t,\mu_f)  \nonumber \\
& +\left(\frac{\alpha_s}{4\pi}\right)^2C^{(2)}_{ij}(s_4,\hat{s},\hat{t}_1,\hat{u}_1,m_t,\mu_f) + \dots\bigg] \, , \label{eq:Cexp}
\end{align}
the $n$-th order term contains a tower of logarithmic plus
distributions, a delta function term, and regular terms in the $s_4\to
0$ limit.  Consider for instance the NNLO coefficient, which is
currently not known.  It has the form
\begin{align}
  \label{eq:C2}
  C^{(2)}(s_4,\hat{s},\hat{t}_1,\hat{u}_1,m_t,\mu) = D_3 \, P_3(s_4) + D_2 \, P_2(s_4)
  + D_1 \, P_1(s_4) + D_0 \, P_0(s_4) + C_0\,\delta(s_4) + R(s_4) \, ,
\end{align}
where the plus distributions 
\begin{align}
  \label{eq:plus}
  P_n(s_4) \equiv \left[ \frac{1}{s_4} \ln^n\frac{s_4}{m_t^2} \right]_+ 
\end{align}
are defined by
\begin{align}
  \int_0^{s_4^{\text{max}}} \left[ \frac{1}{s_4} \ln^n \frac{s_4}{m_t^2}
  \right]_+ g(s_4) = \int_0^{s_4^{\text{max}}} ds_4 \, \frac{1}{s_4} \ln^n\!\bigg(
    \frac{s_4}{m_t^2} \bigg) \left[ g(s_4) - g(0) \right] + \frac{g(0)}{n+1} \ln^{n+1}
  \!\bigg( \frac{s_4^{\text{max}}}{m_t^2} \bigg) \,.
\end{align}
Soft-gluon resummation at NNLL can be used to determine 
the coefficients $D_i$ of the plus-distribution contributions \cite{Kidonakis:2010dk, Ahrens:2011mw}.  The delta-function coefficient,
which is formally of NNNLL order, is unknown, as is the term
$R$, which is non-singular in the $s_4\to 0$ limit and is 
related to hard gluon emission.

In this paper we will discuss the application of the factorization
formula (\ref{eq:fac}) in the soft limit to the high-$p_T$ region of
the double differential cross section, where $m_t \ll p_T$.  Producing
the top quark with high transverse momentum requires that the partonic
center-of-mass energy be large, so in this regime the generic
situation is that $s_4 \ll m_t^2\ll \hat{s},\hat{t}_1,\hat{u}_1$.  We
will refer to such a hierarchy of scales as the double soft and
small-mass limit.  In this limit it is possible to factorize the hard
and soft functions themselves. We explain the form of this
factorization in the remainder of the section.  To simplify the
discussion we ignore for the moment contributions from closed
top-quark loops appearing in virtual corrections.

The factorization of the hard function in the small-mass limit 
was derived in \cite{Ferroglia:2012ku}.  It reads
\begin{align}
\label{eq:hardfact}
 \bm{H}^m_{ij}(\hat{s},\hat{t}_1,\hat{u}_1,m_t,\mu)=
 C_D^2\left(\ln\frac{m_t^2}{\mu^2},\mu\right)  
\bm{H}_{ij}\left(\ln \frac{\hat{s}}{\mu^2}, x_t, \mu\right) +{\cal O}\left(\frac{m_t^2}{\hat{s}}\right)\,.
\end{align}
where $x_t\equiv -\hat{t}_1/\hat{s}$ and we used momentum conservation
$-\hat{u}_1/\hat{s}=1-x_t$ (valid at Born level and also in the soft
limit) in order to simplify dependence on the Mandelstam variables.
This factorization can be thought of as a division of the virtual
corrections into two momentum regions. The hard matrix appearing on
the right-hand side is related to the virtual corrections evaluated
with $m_t=0$ and receives contributions from loop momenta whose
virtuality is at the scale $\hat{s}$, while the coefficient function
$C_D$ contains all the collinear singularities appearing in the limit
$m_t\to 0$ and receives contributions from loop momenta with
virtuality at the scale $m_t^2$. The factorization thus separates
physics from the widely separated scales $m_t^2\ll \hat{s}$.  Two
different ways of deriving (\ref{eq:hardfact}) were discussed in
\cite{Ferroglia:2012ku}.  The first relied on the factorization of the
heavy-quark fragmentation function in the soft limit
\cite{Korchemsky:1992xv, Cacciari:2001cw, Gardi:2005yi,
  Neubert:2007je}, and the second used the factorization formula
\cite{Mitov:2006xs} (see also \cite{Becher:2007cu}) relating massive
amplitudes in the small-mass limit to their massless counterparts.

The factorization of the soft function in the small-mass limit is more
subtle.  Compared to the factorization of the hard function and even
the analogous factorization for the soft function appearing in the
top-pair invariant mass distribution \cite{Ferroglia:2012ku}, a
complication here is that the soft function in the small-mass limit is
characterized by three distinct momentum scales rather than two. In
the next section, we derive the following result:
\begin{align}
 \label{eq:softfact}
  \bm{S}^m_{ij}(s_4,\hat{s},\hat{t}_1,\hat{u}_1,m_t,\mu) &= \int
  d\omega_s \, d\omega_d \, d\omega_b \, \delta(s_4-\omega_s-\omega_d
  -\omega_b) \nonumber
  \\
  &\quad \times \bm{S}_{ij} \left( \omega_s,
    \frac{\omega_s}{\sqrt{\hat{s}}}, x_t, \mu \right) S_D \left(
    \omega_d, \frac{\omega_d m_t}{\hat{s}}, \mu \right) S_B \left(
    \omega_b, \frac{\omega_b}{m_t}, \mu \right) \nonumber
  \\
  &\quad + \mathcal{O}(s_4/m^2_t) + \mathcal{O}(m^2_t/\hat{s}) \, .
\end{align}
At leading order, each of the above functions is a delta function in
its first argument.  At higher orders, the three functions on the
right-hand side are characterized by logarithmic corrections at the
scale shown in their second argument and following from the parametric
relation $\omega_i\sim s_4$.  Before moving on, we discuss the
interpretation of each of these component functions.

First, the massless soft function $\bm{S}_{ij}$ is related to
wide-angle soft real emission corrections to the partonic processes
$(q\bar q,gg)\to Q\bar{Q}$, where $q$ and $Q$ are massless distinct
quarks. Such emissions are associated with a characteristic mass scale
$\mu_s \sim s_4/\sqrt{\hat{s}}$.  This massless soft function is
analogous to that entering the factorization formula for the invariant
mass distribution in the $m_t\to 0 $ limit and calculated to NNLO in
\cite{Ferroglia:2012uy}.  In fact, we will be able to construct
results for the $\bm{S}_{ij}$ to NNLO using calculations from that
paper.

Second, the function $S_D$ describes soft emissions which are
simultaneously collinear to the observed top quark. The
characteristic scale for such soft-collinear emissions is $\mu_d \sim
m_t s_4/\hat{s}$.  This function has to do with the soft part of the
perturbative heavy-quark fragmentation function.  Field theoretically,
it is related to a Wilson-loop operator closing at infinity and containing a finite segment with 
light-like separation. It is equivalent to the partonic shape
function familiar from inclusive $B$ decays, and was 
calculated to NNLO in \cite{Becher:2005pd}.

Finally, the function $S_B$ describes soft emissions which are
simultaneously collinear to the unobserved anti-top quark. The
characteristic scale for this function is $\mu_b \sim s_4/m_t$ (note
that $s_4\ll m_t^2$ is important in this context). Our analysis 
shows that this function is the so-called heavy-quark
jet function introduced in \cite{Fleming:2007qr} and calculated to NNLO in
\cite{Jain:2008gb}. This function is very similar to Wilson-loop
operator used in defining $S_D$, the difference being that it contains
a finite segment with a time-like separation instead of a
light-like one.

By combining (\ref{eq:hardfact}) and (\ref{eq:softfact}) one arrives
at the factorized form of the hard-scattering kernel valid in the
double soft and small-mass limit.  The only subtlety is the treatment 
of terms proportional to $n_h=1$ and related to top-quark loops.
For the counting $s_4\ll m_t^2$, such contributions appear only in
virtual corrections and modify (\ref{eq:hardfact}).  We discuss them
in more detail in Section~\ref{sec:top_loops}.

\section{Factorizing soft real radiation in the small-mass limit}
\label{sec:factorization}

In this section we discuss the factorization of soft real radiation in
the small-mass limit. Such a factorization is equivalent to that of
the massive soft function (\ref{eq:softfact}). We make this clear in
the preliminary discussion below, introducing some notations and
definitions in the process. We then approach the small-mass
factorization in two steps. In Section~\ref{sec:phsp}, we perform a
diagrammatic factorization at NLO using the method of regions. Then,
in Section~\ref{sec:facmasssoft}, we explain how to encode the
all-order contributions from these regions in terms of three distinct
Wilson-loop operators.

The massive soft function can be defined as
\begin{align}
  \label{eq:Sdef1}
  \bm{S}^m(\omega,\hat{s},\hat{t}_1,\hat{u}_1,m_t,\mu) = \frac{1}{d_R}
  \sum_{X} \braket{0|\bm{O}_s^{m\dagger}(0)|X}
  \braket{X|\bm{O}_s^m(0)|0} \, \delta(\omega - 2p_4 \cdot p_{X}) \, ,
\end{align}
where $d_R = N_c$ in the quark annihilation channel and $d_R =
N_c^2-1$ in the gluon fusion channel, with $N_c=3$ colors in QCD. The
final state $X$ is built of soft gluons in the massive theory (i.e.
that relevant for the kinematic limit $s_4 \ll m_t^2,
\hat{s},\hat{t}_1,\hat{u_1}$),\footnote{Here and in the remainder of
  the paper we avoid notational clutter by dropping the hat on the
  partonic state $\hat{X}$.} and
\begin{align}
  \label{eq:WilsonLoop1}
  \bm{O}^m_s(x) = \big[ \bm{S}^m_{v_1} \bm{S}^m_{v_2} \bm{S}^m_{v_3}
  \bm{S}^m_{v_4} \big] (x)
\end{align}
is a Wilson loop operator built out of soft Wilson lines
\begin{align}
  \label{eq:Wilson1}
  \bm{S}^m_{v_i}(x) = \mathcal{P} \exp \left( ig_s \int_{0}^{\infty}
    ds \; v_i \cdot A^a(x+sv_i) \; \bm{T}_i^a \right) ,
\end{align}
where $v_i$ is the velocity vector associated with parton $i$. For
$i=1$, $2$ we have $v_i^2=0$, while for $i=3$, $4$ we have $v_i^2=1$.
We have made use of the basis-independent color-space formalism of
\cite{Catani:1996jh} in our definitions. This allows us to deal
simultaneously with the two cases $(q^{a_1}\bar
q^{a_2},g^{a_1}g^{a_2})\to t^{a_3}\bar t^{a_4}$, where $a_i$ is the
color index of the parton with velocity $v_i$. Details on how to
convert products $\bm{T}_i \cdot \bm{T}_j \equiv \sum_a \bm{T}_i^a
\bm{T}_j^a$ to the basis-dependent matrices used in (\ref{eq:fac}) can
be found, for instance, in \cite{Ferroglia:2012uy}, and we do not
repeat them here. For now we just mention that the amplitude of the
scattering process is represented by an abstract color-space vector
$\ket{\cal M}$, and the generators $\bm{T}^a_i$ act on these vectors
according to rules specific to whether $i$ is a quark or gluon, and in
the initial or final state. For example, the Wilson lines
$\bm{S}_{v_3}$ and $\bm{S}_{v_4}$ in (\ref{eq:Wilson1}) are converted
to
\begin{align}
  \label{eq:Sconvert}
  \bm{S}^m_{v_3}(x) &\to \overline{\mathcal{P}} \exp \left( -ig_s
    \int_{-\infty}^{0} ds \; v_3 \cdot A^a(x+sv_3) \; t^a \right)
  \equiv S_{v_3}^{m\dagger}(x) \, , \nonumber
  \\
  \bm{S}^m_{v_4}(x) &\to \mathcal{P} \exp \left( ig_s
    \int_{-\infty}^{0} ds \; v_4 \cdot A^a(x+sv_4) \; t^a \right)
  \equiv S^m_{v_4}(x) \, ,
\end{align}
where we have used the identification of $-\infty$ and $\infty$ to
bring the definitions of the normal Wilson lines to the convention in
the literature. Other important properties of the color generators are
that $\bm{T}_i\cdot \bm{T}_j = \bm{T}_j\cdot \bm{T}_i$ for $i\neq j$,
and that $\bm{T}_i\cdot \bm{T}_i =C_i$, with $C_i=C_F$ for quarks and
antiquarks and $C_i=C_A$ for gluons. In addition, amplitudes satisfy
color conservation,
\begin{align}
  \label{eq:colorcons}
  \sum_i \bm{T}_i^a \ket{\cal M} = 0 \, .
\end{align}
While this is often expressed as a relation between the generators,
$\sum_i \bm{T}_i^a=0$, it is important to keep in mind that it holds
only when acting on a color-singlet vector, as above.

The soft function takes into account real radiation in the soft limit.
We illustrate this by considering the structure of NLO phase-space
integrals for single-particle inclusive observables. The three-body
phase space for a final state containing the top-quark pair and a
gluon with momentum $k$ is
\begin{multline}
  ({\rm PS})_3 = \int \frac{d^d p_3}{(2\pi)^{d-1}} \, \frac{d^d
    p_4}{(2\pi)^{d-1}} \, \frac{d^d k}{(2\pi)^{d-1}} \, (2\pi)^d \,
  \delta^{(d)}(p_1+p_2-p_3-p_4-k)
  \\
  \times \delta^+(k^2) \, \delta^+(p_3^2-m_t^2) \,
  \delta^+(p_4^2-m_t^2) \, .
\end{multline}
We wish to integrate over the unobserved momenta $p_4$ and $k$. To do
so, we use a technique introduced in \cite{Ellis:1979sj}. The idea is
to shift integration variables to $p_{4k}=p_4+k$ and then split the
phase space up into two Lorentz invariant pieces: that for the
two-to-two production $p_1+p_2\to p_3+p_{4k}$, and that for a
subsequent two-body decay $p_{4k}\to p_4 +k$. We thus write
\begin{multline}
  ({\rm PS})_3 = \frac{1}{(2\pi)^{2d-3}} \int d^dp_3 \, d^d p_{4k} \,
  ds_4 \, \delta^+(p_3^2-m_t^2) \, \delta^+(p_{4k}^2-m_t^2-s_4) \,
  \delta^{(d)}(p_1+p_2-p_3-p_{4k})
  \\
  \times \int d^d k \, d^d p_4 \, \delta^+(k^2) \,
  \delta^+(p_4^2-m_t^2) \, \delta^{(d)}(p_{4k}-p_4-k) \, .
\end{multline}
After a trivial integration the piece on the second line can be written as an integral over the unobserved gluon momentum:
\begin{align}
  \label{eq:phasespace}
  ({\rm PS})_{k} = \int d^d k \, \delta^+(k^2) \,
  \delta^+(s_4-2p_{4k}\cdot k) \, .
\end{align}
The piece on the first line can be arranged into a form appropriate
for describing the double differential 1PI observables and is
unimportant for what follows.

To evaluate the NLO real emission corrections to the differential
cross section one integrates the squared matrix element over the phase
space (\ref{eq:phasespace}). The structure of these phase-space
integrals simplifies in the soft limit $k\to 0$, in which case $s_4
\ll m_t^2, \hat{s},\hat{t}_1,\hat{u}_1$. In this limit one can replace
the squared matrix element by eikonal factors for a gluon emission
from each leg, approximate $p_{4k}\sim p_4$ in the delta-function
constraint, and drop any $k$ dependence in the matrix element arising
from the shift $p_4\to p_{4k}-k$. One must then evaluate integrals of
the form\footnote{The integrals $I_{ij}$ in (\ref{eq:softints}) are
  connected to the position space integrals ${\mathcal I}'_{ij}$ in
  Eq.~(20) of Ref.~\cite{Ahrens:2011mw} through relation
  \begin{align*}
    {\mathcal I}'_{ij}(x_0) = -2 \, (4 \pi e^{-\gamma_E})^\epsilon
    \int_0^\infty d s_4 \, \exp \left(-\frac{s_4}{m_t
        e^{\gamma_{\mbox{{\tiny E}}}} \mu e ^{(-L_0/2)}}\right)
    I_{ij}^m(s_4) \, .
  \end{align*}
}
\begin{align}
  \label{eq:softints}
  I^m_{ij} &= \pi^{-1+\epsilon} e^{\epsilon\gamma_E} \mu^{2\epsilon}
  \int d^dk  \delta^+(k^2) \, \delta^+(s_4-2p_{4k}\cdot k) \, \frac{v_i\cdot v_j}{v_i
    \cdot k\,v_j \cdot k} \nonumber
  \\
  &\equiv \int [dk] \, \delta^+(s_4-2p_{4k}\cdot k) \, \frac{v_i\cdot
    v_j}{v_i \cdot k\,v_j \cdot k} \, .
\end{align}  
We have introduced factors convenient for the $\overline{\text{MS}}$
renormalization scheme,  and absorbed them into the integral measure
$[dk]$ defined on the second line. The quantity  $\epsilon =
(4-d)/2$ is the dimensional regulator.  These integrals are exactly those
appearing in the NLO corrections to the soft function
(\ref{eq:Sdef1}), which shows explicitly its connection with real
radiation. In fact, the NLO bare soft function is calculated by
associating a color factor $\bm{T}_i\cdot \bm{T}_j$ with each integral
and summing over possible attachments to the partons $i$, $j$. A first
step to factorizing the soft function in the small-mass limit is thus
to understand the structure of the integrals (\ref{eq:softints}). We
turn to this problem in the following subsection, using the method of
regions as a tool for performing a diagrammatic factorization.

We end this section with some comments concerning the arguments of the
massive soft function (\ref{eq:Sdef1}). The Wilson lines entering its
definition depend on the velocity vectors $v_i$, so the object on the
left-hand side depends on invariants formed from the velocities and
$p_4 =m_t v_4$. In order to keep contact with our physical picture of
the soft function as representing soft real radiation, we express
these scalar products in terms of the Mandelstam variables. However,
in studying the properties of the integrals it is sometimes useful to
keep the structure of the scalar products explicit. For instance, by
considering properties of the integrals (\ref{eq:softints}) under
simultaneous rescalings of the different vectors and $s_4$ (see, for
instance, \cite{Becher:2012za}) one finds their general functional
form is
\begin{align}
  \label{eq:scalingint}
  I_{ij}^m = \frac{1}{s_4} \, f\left(s_4 \sqrt{\frac{v_i\cdot v_j}{p_4
        \cdot v_i \, p_4 \cdot v_j}}\right) .
\end{align}

\subsection{NLO phase space integrals and momentum regions}
\label{sec:phsp}

The NLO integrals (\ref{eq:softints}) were evaluated for arbitrary
$m_t$ in \cite{Ahrens:2011mw}. Here we are interested in the
asymptotic expansion of those integrals in the small-mass limit, where
$m_t^2\ll \hat{s},\hat{t_1},\hat{u}_1$. To leading order in
$m_t^2/\hat{s}$ the results are
\begin{align}
  \label{eq:Iij}
  I^m_{12} &= \frac{1}{s_4}
  \left(\frac{s_4}{\sqrt{\hat{s}}\mu}\right)^{-2\epsilon}
  (x_t\bar{x_t})^{\epsilon} \left( -\frac{2}{\epsilon} +
    \frac{\pi^2}{6}\epsilon \right) , \nonumber
  \\
  I^m_{13} &= \frac{1}{s_4}
  \left(\frac{s_4}{\sqrt{\hat{s}}\mu}\right)^{-2\epsilon} \left(
    -\frac{1}{\epsilon} + \ln\left(\frac{\hat{s}}{m_t^2}\right) +
    2\ln\left(\frac{x_t}{\bar{x_t}} \right) + \frac{\epsilon}{2}
    \left[ \ln^2\left(\frac{\hat{s}}{m_t^2}\right) -
      2\ln^2\left(\frac{x_t}{\bar{x_t}} \right) + \frac{\pi^2}{2}
    \right] \right) , \nonumber
  \\
  I^m_{23} &= I^m_{13}(x_t \leftrightarrow \bar{x_t}) \, , \nonumber
  \\
  I^m_{33} &= \frac{2}{s_4}
  \left(\frac{m_t}{\mu}\frac{s_4}{\hat{s}}\right)^{-2\epsilon} \, ,
  \nonumber
  \\
  I^m_{14} &= I^m_{24} = \frac{1}{s_4}
  \left(\frac{s_4}{m_t\mu}\right)^{-2\epsilon} \left(
    -\frac{1}{\epsilon} + \frac{\pi^2}{4} \epsilon \right) , \nonumber
  \\
  I^m_{34} &= \frac{1}{s_4}
  \left(\frac{s_4}{\sqrt{\hat{s}}\mu}\right)^{-2\epsilon} \left(
    2\ln\left(\frac{\hat{s}}{m_t^2}\right) + \frac{\pi^2}{3}\epsilon
  \right) , \nonumber
  \\
  I^m_{44} &= \frac{1}{s_4} \left(\frac{s_4}{m_t\mu}\right)^{-2\epsilon} (2+4\epsilon) \, .
\end{align}
In the above equations, we have defined $\bar x_t = 1-x_t$. These explicit results make
clear that some of the integrals are characterized by a single mass
scale, while some of them depend on more than one mass scale and
contain logarithms of $m_t^2/\hat{s}$.

We will now show how to reproduce these results using the method of
regions \cite{Beneke:1997zp}. This allows us to factorize the
multiscale integrals into a sum of simpler, one-scale integrals. While
this method was originally developed to construct the asymptotic
expansions of loop integrals and is usually discussed in that context,
it applies equally well to the phase-space integrals considered here.
At the technical level, the reason for this is that integrals such as
(\ref{eq:softints}) are equivalent to loop integrals, since one can
rewrite the delta-function constraint as the discontinuity of
propagators (see for example \cite{Anastasiou:2002yz}). Rather than
actually doing this, one can simply apply the normal procedure for
expanding loop diagrams by regions to the phase-space integrals
directly. This proceeds as follows. First, one defines a region by
associating a specific scaling to the components of the undetermined
momentum $k$ in terms of the external expansion parameter (in our case
$m_t^2/\hat{s}$) . One then expands the integrand as appropriate for
the particular momentum region, and integrates over the whole phase
space. After finding all of the possible momentum regions which
contribute at a given power, one adds their contributions together to
obtain the asymptotic expansion of the full integral.

The exact scalings of the regions which contribute to the integrals
(\ref{eq:softints}) in the small-mass limit are perhaps not obvious at
first sight.  However, physical intuition suggests three
possibilities: wide-angle soft emission, soft emission collinear to
the observed top quark, and soft emission collinear to the unobserved
anti-top quark.  The regions analysis below shows that this is 
indeed correct, and moreover fixes the momentum scale associated
with each of these regions.  

To discuss the momentum regions, let us first introduce four
light-like vectors $n_1$, $n_2$, $n_3$ and $n_4$, whose space
components are aligned with the momenta $p_1$, $p_2$, $p_3$ and
$p_{4k}$, respectively. For convenience we normalize the vectors to
satisfy $n_1 \cdot n_2 = n_3 \cdot n_4 = 2$. The other scalar products
are then fixed to $n_1 \cdot n_3 = n_2 \cdot n_4 = 2x_t$ and $n_1
\cdot n_4 = n_2 \cdot n_3 = 2\bar{x}_t$. Picking two reference vectors
$n_i$ and $n_j$, we define the light-cone decomposition of an
arbitrary four-vector $k$ as
\begin{align}
  k^\mu &= k_{+ij} \, \frac{n_i^\mu}{\sqrt{2 \, n_i \cdot n_j}} +
  k_{-ij} \, \frac{n_j^\mu}{\sqrt{2 \, n_i \cdot n_j}} + k_{\perp
    ij}^\mu \, , \nonumber
  \\
  k_{+ij} & = \frac{n_j \cdot k}{\sqrt{n_i \cdot n_j / 2}} \, , \quad
  k_{-ij} = \frac{n_i \cdot k}{\sqrt{n_i \cdot n_j / 2}} \, , \quad
  k_{Tij}^2 = -k_{\perp ij}^2 \, . \label{eq:lightcone}
\end{align}
In the following we drop the $ij$ labels when there is no danger of
confusion. A judicious choice of the light-cone vectors for a given
integral can significantly simplify calculations, as will become
evident in the examples below. For the discussion of regions, it is
particularly convenient to choose $i=3$ and $j=4$. The scaling of the
momentum $p_{4k}$ in the limit $s_4 \ll m_t^2 \ll \hat{s}$ is then
given by
\begin{align}
  p_{4k}^\mu = (p_{4k+},p_{4k-},p_{4k\perp}) \sim \sqrt{\hat{s}} (\lambda^2,1,0) \, ,
\end{align}
where $\lambda=m_t/\sqrt{\hat{s}}$. The delta function in
(\ref{eq:softints})  constrains the components of $k$ to satisfy
\begin{align}
  \lambda^2 k_- + k_+ \sim \sqrt{\hat{s}} \, \frac{s_4}{m_t^2} \,
  \lambda^2 \, .
\end{align}
We can use the relations
\begin{align} \label{scaling-velocities}
  v_1^\mu \sim v_2^\mu \sim (1,1,1) \, , \quad v_3^\mu \sim
  \frac{1}{\lambda} (1,\lambda^2,0) \, , \quad v_4^\mu \sim
  \frac{1}{\lambda} (\lambda^2,1,0) \, ,
\end{align}
to analyze the leading behavior of the propagators $1/(v_i \cdot k)$
and $1/(v_j \cdot k)$.  This power-counting exercise is sufficient 
to identify the three relevant momentum regions:
\begin{subequations}   \label{eq:scalingmom1}
\begin{alignat}{2}
k^{\mu}_s &\sim \frac{s_4}{\sqrt{\hat{s}}} \sim \sqrt{\hat{s}} \,
  \frac{s_4}{m_t^2} \, (\lambda^2,\lambda^2,\lambda^2) &\qquad&
  \text{(soft, wide angle)}, \label{scaling-soft}
  \\
  k^{\mu}_{sc} &\sim \frac{s_4 \, p_3^\mu}{\hat{s}} \sim
  \sqrt{\hat{s}} \, \frac{s_4}{m_t^2} \,
  (\lambda^2,\lambda^4,\lambda^3) && \text{(soft, collinear to the
    top)}, \label{scaling-sc}
  \\
  k^{\mu}_{sc'} &\sim \frac{s_4 \, p_4^\mu}{m_t^2} \sim \sqrt{\hat{s}}
  \, \frac{s_4}{m_t^2} \, (\lambda^2,1,\lambda) && \text{(soft,
    collinear to the anti-top)}. \label{scaling-scp}
\end{alignat}
\end{subequations}
However, not every region contributes to each integral. For example,
it is clear from power-counting that the $sc$ (i.e.\ soft, collinear to the top) region 
only contributes
to integrals involving $v_3$, while the $sc'$ (i.e.\ soft, collinear to the anti-top) region only contributes
to integrals involving $v_4$. In the following, we structure our
discussion by analyzing how the three regions contribute to the list
of integrals in (\ref{eq:Iij}).

\paragraph{Wide-angle soft emission.}

We first discuss the wide-angle soft region. In this region, to
leading power in $\lambda$, we can approximate $2p_{4k} \cdot k \approx
\sqrt{\hat{s}} n_4 \cdot k$ in the delta function and also
\begin{align}
  \frac{v_i \cdot v_j}{v_i \cdot k \; v_j \cdot k} \approx \frac{n_i
    \cdot n_j}{n_i \cdot k \; n_j \cdot k}
\end{align}
for the propagators. The contribution to the integral $I^m_{ij}$ from
the soft region is then given by the integral
\begin{align}
  I^s_{ij} = \int [dk_s] \, \delta^+(s_4-\sqrt{\hat{s}} n_4 \cdot k_s)
  \, \frac{n_i \cdot n_j}{n_i \cdot k_s \; n_j \cdot k_s} \, .
\end{align}
Note that the factor of $\sqrt{\hat{s}}$ in the definition of the
wide-angle soft region is a necessary condition for the delta function
constraint to be satisfied, and in fact explains why this particular
scaling appears. 

The above integral is straightforward to evaluate.\footnote{A
  step-by-step derivation is given in \cite{Becher:2012za}.}  It is
instructive to write the result in the following way:
\begin{align}
  \label{eq:I12s}
  I_{ij}^s &= \frac{1}{s_4}
  \left(\frac{s_4}{\sqrt{\hat{s}}\mu}\right)^{-2\epsilon} \left(
    \frac{2 n_i \cdot n_j}{n_4 \cdot n_i \, n_4 \cdot n_j}
  \right)^{-\epsilon} \left( -\frac{2}{\epsilon} +
    \frac{\pi^2}{6}\epsilon \right) .
\end{align}
The above result has several important features. First, the mass scale
on which it depends is characterized by $\mu_s \sim
s_4/\sqrt{\hat{s}}\sim |k_s|$. The scaling of $k_s$ enforced under the
integrand determines the mass scale in the integral. Second, the
dependence on the light-cone vectors $n_i$ is of the form required by
(\ref{eq:scalingint}).  Finally, the result is non-zero only if $i\neq
j$, and if $i,j\neq 4$, because otherwise one of the scalar
products in (\ref{eq:I12s}) vanishes and the prefactor is zero in
dimensional regularization. We can see this also in intermediate
results. An explicit example is the following integral:
\begin{align}
  I^{s}_{14} &= \int [dk_s] \, \delta^+(s_4- \sqrt{\hat{s}} n_4 \cdot
  k_s) \, \frac{n_1 \cdot n_4}{n_1 \cdot k_s \; n_4 \cdot k_s}
  \nonumber
  \\
  &= \int [dk_s] \, \delta^+(s_4- \sqrt{\hat s} n_4\cdot k_s) \,
  \frac{n_1 \cdot n_4}{n_1 \cdot k_s} \, \frac{\sqrt{\hat{s}}}{s_4} =
  0 \, .
\end{align}
The equality follows because the integral is scaleless, as one can
verify by choosing $n_1$ and $n_4$ as basis vectors for the light-cone
coordinate decomposition (\ref{eq:lightcone}) and then integrating
over the transverse and $k_s\cdot n_4$ components. The reason this
happens is that when a gluon connects partons $i$ and $j$, the more
precise definition of wide-angle soft is
\begin{align}
  k_s^{ij} \sim \frac{s_4}{\sqrt{\hat{s}}} \, (n_i + n_j) \, .
\end{align}      
If, say, $n_j=n_4$, one must drop its contribution inside the
delta-function constraint, in which case the square of the soft
momentum vanishes and the integral is scaleless.

The total contribution from the wide-angle soft region to the NLO
phase-space integrals is obtained by associating a factor of 
$\bm{T}_i\cdot \bm{T}_j$ with each integral $I^s_{ij}$ and summing 
over legs.  The result is proportional to 
\begin{align}
  I^{s} &=2 \bm{T}_1 \cdot \bm{T}_2 \, I_{12}^{s}
+2 \bm{T}_1 \cdot \bm{T}_3 \, I_{13}^{s}+
2 \bm{T}_2 \cdot \bm{T}_3 \, I_{12}^{s} \, . \label{eq:sregion}
\end{align}
The  contributions above are
derived from the general integrals after the replacement $v_i \to
n_i$. The time-like vectors are expanded out into light-like ones,
which corresponds to calculating real emission corrections with
massless partons. We use this fact in the next section to define the
``massless" soft function, and calculate it to NNLO in
Appendix~\ref{sec:SoftFunction}.  The NLO result for the bare soft
function is exactly that given in (\ref{eq:sregion}), showing 
the direct correspondence between the operator definition and
regions calculation.

\paragraph{Soft emission collinear to the top quark.}

We next consider soft emission which is simultaneously collinear to
the top quark. We call this region soft-collinear or simply $sc$. The
scaling of soft-collinear momenta is $k_{sc}^\mu \sim p_3^\mu
s_4/\hat{s}$. In contrast to the wide-angle soft region, to expand the
integrand in the soft-collinear region we must keep the
$m_t$-dependence in the parameterization of $p_3$, i.e, $p_3^\mu =
\sqrt{\hat{s}}n_3^\mu/2 + n_4^\mu m_t^2/2\sqrt{\hat{s}}$, such that
$v_3^2=p_3^2/m_t^2=1$.
For all other velocities $v_i$ with $i\neq 3$, it is enough to know
that $v_i \cdot v_3 \approx v_i^-v_3^+/2$ and $v_i \cdot k_{sc}
\approx v_i^-k_{sc}^+/2$, no further specifications are needed.

We can now consider contributions from the soft-collinear region,
starting with that to $I_{13}^m$. Using the scalings in (\ref{scaling-velocities}) and (\ref{scaling-sc}) to perform
the expansion under the integrand, we find to leading order in
$\lambda$:
\begin{align}
  I^{sc}_{13} &= \int [dk] \, \delta^+(s_4-\sqrt{\hat s} k_{sc}^+) \,
  \frac{2 v_3^+ }{(v_3^+ k_{sc}^- +v_3^- k_{sc}^+) \; k_{sc}^+}
  \nonumber
  \\
  &= \frac{1}{s_4} \left(\frac{s_4m_t}{\hat{s}\mu}\right)^{-2\epsilon}
  \left( \frac{1}{\epsilon} + \frac{\pi^2}{12}\epsilon \right) ,
\end{align}  
where the second equality follows after a straightforward integration.
Note that the integral is characterized by the single mass scale
$\mu^2 \sim k_{sc}^2$, and that the scaling of $k_{sc}^+$ is such that
the two terms in the delta-function are of the same order. Moreover,
the integral contains no information about the velocity $v_1$. It is
therefore easy to show that $I^{sc}_{13}=I^{sc}_{23}=I^{sc}_{34}$. The
only other contribution from the soft-collinear region is to
$I_{33}^m$, which is in fact  saturated by that region:
\begin{align}
  I^{sc}_{33}=\int [dk_{sc}] \, \delta^+(s_4-\sqrt{\hat s} k_{sc}^+)
  \, \frac{1}{v_3 \cdot k_{sc}\, v_3 \cdot k_{sc}} = I^m_{33} \, .
\end{align}   
It is obvious that the soft-collinear region contributes only to
integrals involving $v_3$, so this completes the analysis.

The total contribution from the soft-collinear region to the massive
soft function is obtained by associating a factor of $\bm{T}_i\cdot
\bm{T}_j$ with each integral $I^{sc}_{ij}$ and summing over legs. The
result is proportional to
\begin{align}
  I^{sc} &=\bm{T}_3 \cdot \bm{T}_3 \, I_{33}^{sc}+ 2 I_{13}^{sc}
  \sum_{i\neq 3}\bm{T}_i\cdot \bm{T}_3 = C_F (I_{33}^{sc} -2 I
  _{13}^{sc}) \, , \label{eq:scregion}
\end{align}
In the second equality we used color conservation
(\ref{eq:colorcons}), after which one sees that the contribution is
diagonal in color space. Furthermore, the expansion in the
soft-collinear region is such that the delta-function constraint has
the form $\delta(s_4 - \sqrt{\hat{s}}n_4 \cdot k_{sc})$, i.e. the
constraint vector $n_4$ is light-like. We will see in the next section
that both of these features are important when identifying the
contributions of this region with the soft part of the heavy-quark
fragmentation function $S_D$ defined in (\ref{eq:SDdef2}) below. In
fact, one can check that the NLO bare contributions to that function
are exactly reproduced by (\ref{eq:scregion}), which is especially
obvious after writing down the NLO integrals using the Feynman rules
for Wilson lines and noting the correspondence with the integrands
expanded in the soft-collinear region.

\paragraph{Soft emission collinear to the anti-top quark.}

Finally, we consider soft emission which is simultaneously collinear
to the unobserved anti-top quark. The scaling of such $sc'$ momenta is
$k_{sc'}^\mu \sim s_4 p_4^\mu/m_t^2$. To perform an expansion in this
region we parametrize $p_4 = \sqrt{\hat{s}}n_4/2
+ n_3m_t^2/2\sqrt{\hat{s}}$, and then $v_4^2=p_4^2/m_t^2=1$. 
For all other velocities $v_i$ with $i\neq 4$, we can approximate $v_i
\cdot v_4 \approx v_i^+v_4^-/2$ and $v_i \cdot k_{sc'} \approx
v_i^+k_{sc'}^-/2$.

The analysis of the contributions from the $sc'$ region to the
integrals is very similar to that of the $sc$ region.  Using the scalings (\ref{scaling-velocities}) and (\ref{scaling-scp}), the contribution
to $I_{14}^m$ from this region is:
\begin{align}
  I^{sc'}_{14} &= \int [dk_{sc'}] \, \delta^+(s_4 - 2m_t v_4 \cdot
  k_{sc'}) \, \frac{v_4^-}{v_4 \cdot k_{sc'} \; k_{sc'}^-} \nonumber
  \\
  &= \frac{1}{s_4} \left(\frac{s_4}{m_t \mu}\right)^{-2\epsilon}
  \left( -\frac{1}{\epsilon} + \frac{\pi^2}{4} \epsilon \right) .
\end{align}  
The final equality follows from direct integration using the standard
techniques. One sees that $I_{14}^{sc'}=I_{14}^m$. The integrals show
familiar features: the particular scaling of $k_{sc'}$ (\ref{scaling-scp}) ensures
that the two terms in the argument of the delta-function scale the
same, and the characteristic scale is $\mu^2 \sim k_{sc'}^2$.
Moreover, the integral depends only on quantities related to parton 4.
One can show that $I^{sc'}_{24}= I^{sc'}_{34}= I^{sc'}_{14}$.
Furthermore, $I^{sc'}_{44}=I^{m}_{44}$. The total contribution of this
region to the NLO soft function is thus proportional to
\begin{align}
  I^{sc'} &=\bm{T}_4 \cdot \bm{T_4} \, I_{44}^{sc'} + 2 I_{14}^{sc'}
  \sum_{i\neq 4} \bm{T}_i \cdot \bm{T}_4 = C_F (I_{44}^{sc'} - 2 I
  _{14}^{sc'}) \, . \label{eq:scpregion}
\end{align}   
As was the case with the emissions collinear to the top, the total
contribution is color diagonal. The two regions do not, however, give
identical contributions. The reason for this is that while after
expansion in the $sc$ region the delta-function constraint involves a
light-like vector $n_4$, in the $sc'$ region the delta-function
constraint involves the time-like vector $v_4$. Instead of being
related to the heavy-quark fragmentation function, these contributions
are related to a different object, the heavy-quark jet function $S_B$
defined in (\ref{eq:Bdef}) below. Here again one can check that the
NLO contributions to $S_B$ are exactly those arising from
(\ref{eq:scpregion}).

\paragraph{Comments.}

To summarize, we have found three distinct momentum regions: soft,
associated with the scale $\mu_s \sim s_4/\sqrts$; soft and collinear
to $p_3$, associated with the scale $\mu_{sc}\sim m_t s_4/\hat{s}$;
and soft and collinear to $p_4$, associated with the scale
$\mu_{sc'}\sim s_4/m_t$. Although all of these scales vanish in the
limit $s_4 \to 0$, the method of regions provides a technical way of
separating out their contributions to the soft function, and
identifying the exact mass scale associated with them.

One could use the regions method to prove the factorization formula
(\ref{eq:softfact}) to all orders  diagrammatically.  
It is convenient instead to use effective field theory to reorganize
contributions from the different regions into field-theoretical
objects encoding their all-order structure, a problem we turn to next.
In either case, one might wonder if the three regions identified here
are sufficient also at higher orders.  Our explicit checks on
factorization described below have shown that this is the case at
least to NNLO.  We have no proof beyond that, yet also see no physical
effect (other than complications from heavy-quark loops we deal with
later) that would give rise to other regions.  This is an assumption
in the ``all-order" analysis that follows, and is common to most
``proofs" of factorization relying on the regions method, effective
field-theory based or not.

\subsection{All-order factorization in the small-mass limit}
\label{sec:facmasssoft}

Having identified the momentum regions which contribute to the
phase-space integrals in the double soft and small-mass limit, we are
now in position to explore their all-order structure. There are two
possible routes to doing so. The first is to construct an appropriate
version of soft-collinear effective theory and apply it to double
differential cross sections for 1PI observables using a multistep
matching procedure. Many of the steps of such a construction can be
taken over from \cite{Ahrens:2010zv}, for the soft limit, and from
\cite{Fleming:2007qr}, for the boosted limit.  A second, more direct
route is to start from the definition of the soft function
(\ref{eq:Sdef1}) for arbitrary $m_t$, factorize the QCD gluon field
appearing in the single Wilson-loop operator into a sum of fields
whose Fourier components are restricted to certain regions, and then
see how the different component fields factorize into operators. We
pursue this second method here, and then comment on the alternate
derivation at the end of the section.

Our aim is to decompose the operator definition of the massive soft
function (\ref{eq:Sdef1}) into component operators whose diagrammatic
expansions encode the contributions of the three distinct momentum
regions. These operators are functions of gluon fields whose Fourier
components are restricted to the scalings appropriate for a particular
region. We write the decomposition as
\begin{align} 
  A^a \to A^a_s + A^a_{sc} +A^a_{sc'} \, . \label{eq:split}
\end{align}
The Wilson lines in the definition (\ref{eq:Wilson1}) then decompose
into a product of three Wilson lines containing gluons of the
different scalings. This works as follows. Let us first
define
\begin{align}
  \bm{S}_{v_i}(x) &= \mathcal{P} \exp \left( ig_s \int_0^{\infty} ds
    \; v_i \cdot A^a_s(x+sv_i) \, \bm{T}^a_i \right) , \nonumber
  \\
  \bm{Y}_{v_i}(x) &= \mathcal{P} \exp \left( ig_s \int_0^{\infty} ds
    \; v_i \cdot A^a_{sc}(x+sv_i) \, \bm{T}^a_i \right) , \nonumber
  \\
  \bm{Y}'_{v_i}(x) &= \mathcal{P} \exp \left( ig_s \int_0^{\infty} ds
    \; v_i \cdot A^a_{sc'}(x+sv_i) \, \bm{T}^a_i \right) \, .
\end{align}
By using the following identity for path-ordered exponentials:
\begin{multline}
  \mathcal{P} \exp \left[ \int_a^b dx \left( A(x) + B(x) \right)
  \right]
  \\
  = \mathcal{P} \exp \left[ \int_a^b dx \, A(x) \right] \; \mathcal{P}
  \exp \left[ \int_a^b dx \left( \mathcal{P} e^{\int_a^x dx' A(x')}
    \right)^{-1} B(x) \left( \mathcal{P} e^{\int_a^x dx' A(x')}
    \right) \right] ,
\end{multline}
it is easy to show that
\begin{align}
  \label{wilson-fact}
  \bm{S}^m_{v_i}(x) = \bm{Y}_{v_i}(x) \, \bm{\tilde{S}}_{v_i}(x) \,
  \tilde{\bm{Y}}'_{v_i}(x) \, ,
\end{align}
where
\begin{align}
  \label{eq:tildeWilson}
  \tilde{\bm{S}}_{v_i}(x) &= \mathcal{P} \exp \left( ig_s
    \int_0^{\infty} ds \left[ v_i \cdot A^a_s \,
      \bm{Y}_{v_i}^{\dagger} \bm{T}^a_i \bm{Y}_{v_i} \right] (x+sv_i)
  \right) , \nonumber
  \\
  \tilde{\bm{Y}}'_{v_i}(x) &= \mathcal{P} \exp \left( ig_s
    \int_0^{\infty} ds \left[ v_i \cdot A^a_{sc'} \,
      \tilde{\bm{S}}_{v_i}^{\dagger} \bm{Y}_{v_i}^{\dagger} \bm{T}^a_i
      \bm{Y}_{v_i} \tilde{\bm{S}}_{v_i}\right] (x+sv_i) \right) .
\end{align}
We now need to perform a consistent power expansion in $\lambda$. This
involves the expansion of the gluon fields themselves as well as their
momenta. In the soft-collinear effective theory, the gluon fields
scale the same as their momenta \cite{Bauer:2000yr, Beneke:2002ph}. It
is then clear that only the `+' component of the $A_{sc}$ field and
its momentum needs to be kept when it interacts with the $A_s$ or the
$A_{sc'}$ field. The same is true for the $A_{s}$ field when it
interacts with the $A_{sc'}$ field. This is often called ``multipole
expansion'' in the literature \cite{Beneke:2002ph}. After this
expansion, the velocity vectors in the Wilson lines $\bm{Y}_{v_i}$ and
$\tilde{\bm{S}}_{v_i}$ on the right side of (\ref{eq:tildeWilson}) can
be replaced by their components along the plus direction, which in our
reference system of choice is $n_4$, i.e.  $v_i \to n_4$
in~(\ref{eq:tildeWilson}). We then redefine the fields as
\begin{align}
  \label{eq:field-red}
  A^a_s(x) \, \bm{Y}_{n_4,i}^{\dagger}(x_-) \bm{T}^a_i
  \bm{Y}_{n_4,i}(x_-) &\to A^a_s(x) \, \bm{T}^a_i \, , \nonumber
  \\
  A^a_{sc'}(x) \, \tilde{\bm{S}}_{n_4,i}^{\dagger}(x_-)
  \bm{Y}_{n_4,i}^{\dagger}(x_-) \bm{T}^a_i \bm{Y}_{n_4,i}(x_-)
  \tilde{\bm{S}}_{n_4,i}(x_-) &\to A^a_{sc'}(x) \, \bm{T}^a_i \, ,
\end{align}
where 
\begin{align}
  \bm{Y}_{n_4,i}(x_-) &= \mathcal{P} \exp \left( ig_s \int_0^{\infty}
    ds \; n_4 \cdot A^a_{sc}(x_- +sn_4) \, \bm{T}^a_i \right)
\end{align}
is a Wilson line along the direction of $n_4$ but in the color
representation of parton $i$, and similarly for
$\tilde{\bm{S}}_{n_4,i}$. To show that these field redefinitions are
actually the same for each $i$, we use the identity 
\begin{align}
  \bm{Y}_{n_4,i} \bm{T}^a_i \bm{Y}^\dagger_{n_4,i} =Y_{n_4,\text{adj}}^{ba} \,
  \bm{T}^b_i \, ,
\end{align}
which holds whether $\bm{T}_i$ is the color generator for a quark
or a gluon. Here $Y_{n_4,\text{adj}}$ is a Wilson line in the adjoint
representation along the direction of $n_4$, 
\begin{align}
  Y^{ab}_{n_4,\text{adj}}(x) = \mathcal{P} \exp \left[ ig_s
    \int_0^{\infty} ds \; n_4 \cdot A^c_{sc}(x+sn_4) \,
    T^c_{\text{adj}} \right]^{ab} \, ,
\end{align}
with $(T^c_{\text{adj}})^{ab}=-if^{cab}$. It then follows that the
field redefinition of, e.g. the soft gluon field is
\begin{align}
  \label{eq:field-red-2}
  \left[ Y^\dagger_{n_4,\text{adj}}(x_-) \right]^{ab} A^b_s(x) \to
  A^a_s(x) \, ,
\end{align}
which does not involve the particular color generator at all.

After the field redefinitions, the two functions in
(\ref{eq:tildeWilson}) are just usual Wilson lines in terms of the new
fields. We work with the redefined  fields in the following, and
drop the tilde on the two functions. In soft-collinear effective
theory, such  field redefinitions also remove the interactions among
the various fields in the Lagrangian and are therefore referred to as
the ``decoupling transformations'' \cite{Bauer:2001yt}. It is 
clear from (\ref{eq:field-red-2}) that the field redefinitions on the
gluon fields made above are equivalent to those in
\cite{Bauer:2001yt}.  

We have now achieved a decomposition of the original Wilson lines into
three separate Wilson lines, each involving gluon fields with a
particular scaling. Moreover, the gluon fields with different scalings
no longer interact. To factorize the matrix element, we then use that
the inclusive state $X$ can be written as a product of states
involving $s$, $sc$ and $sc'$ gluons. Moreover, the $2p_4 \cdot p_X$
factor in the delta function can be written as a sum of the
contributions from these modes. We use these facts to write
\begin{align}
  \label{eq:Smfact1}
  \bm{S}^m(s_4,\hat{s},\hat{t}_1,m_t,\mu) &= \int d\omega_s \,
  d\omega_{sc} \, d\omega_{sc'} \, \delta(s_4-\omega_s
  -\omega_{sc}-\omega_{sc'}) \nonumber
  \\
  &\times \frac{1}{d_R} \sum_{X_s} \braket{0|\bm{O}_s^\dagger(0)|X_s}
  \braket{X_s|\bm{O}_s(0)|0} \, \delta(\omega_s - 2 p_4 \cdot p_{X_s})
  \nonumber
  \\
  &\times \sum_{X_{sc}} \braket{0|O_{sc}^\dagger(0)|X_{sc}}
  \braket{X_{sc}|O_{sc}(0)|0} \, \delta(\omega_{sc} - 2 p_4 \cdot
  p_{X_{sc}}) \nonumber
  \\
  &\times \sum_{X_{sc'}} \braket{0|O_{sc'}^\dagger(0)|X_{sc'}}
  \braket{X_{sc'}|O_{sc'}(0)|0} \, \delta(\omega_{sc'} - 2 p_4\cdot
  p_{X_{sc'}}) \nonumber
  \\
  &+ {\cal O}\left(\frac{m_t^2}{\hat{s}}\right) + {\cal
    O}\left(\frac{s_4}{m_t^2}\right) .
\end{align}
The above Wilson line operators (defined in (\ref{eq:Sdef}),
(\ref{eq:SDdef-cntd}), and (\ref{eq:SBdef-cntd}) below) arise after a
multipole expansion appropriate for the particular momentum region,
according to the rules which we explain below. This expansion ensures
that the Feynman rules for the Wilson-line attachments in the
different regions are such that they produce the correct, homogeneous
expansion appropriate for the momentum region the gluon fields are
restricted to. The form of this expansion is very similar to what
appeared in the regions analysis in the previous subsection. We thus
structure our discussion in a similar way, performing a
region-by-region analysis which leads to operator definitions of the
objects in (\ref{eq:Smfact1}).

\paragraph{The wide-angle soft region.}
Let us first consider the wide-angle soft region. We parametrize the
external momenta as described in the previous subsection. The expansion
inside the delta-function and Wilson lines then reads
\begin{align}
  \delta(\omega_s - 2p_4 \cdot p_{X_s}) &\approx \delta(\omega_s -
  \sqrts \, n_4 \cdot p_{X_s}) \, , \nonumber
  \\
  v_i \cdot A^a_s(sv_i) &\to n_i \cdot A^a_s(sn_i) \, .
\end{align}
Note that for $i=3,4$, we have $v_i^\mu \approx (\sqrts/2m_t)
n_i^\mu$. However, an important property of Wilson lines is their
invariance under rescalings of the reference vector $n_i\to \lambda
n_i$, for an arbitrary number $\lambda$, which can be verified immediately
from the definition (\ref{eq:Wilson1}) after a change of variables. We
used this fact to eliminate factors of $\sqrts/2m_t$. The expansion
above implies that in the wide-angle soft region we can treat all
partons as massless, replacing  the time-like vectors
$v_3$ and $v_4$ with light-like ones $n_3$ and $n_4$.

From the above discussion, we are led to define the massless soft
function as
\begin{align}
  \label{eq:Sdef}
  \bm{S} \left( \omega_s, \frac{\omega_s}{\sqrt{\hat{s}}}, x_t, \mu
  \right) = \frac{1}{d_R} \sum_{X_s}
  \braket{0|\bm{O}_s^\dagger(0)|X_s} \braket{X_s|\bm{O}_s(0)|0} \,
  \delta(\omega_s - \sqrts \, n_4\cdot p_{X_s}) \, ,
\end{align}
where 
\begin{align}
  \label{eq:WilsonLoop}
  \bm{O}_s(x) = \big[ \bm{S}_{n_1} \bm{S}_{n_2} \bm{S}_{n_3}
  \bm{S}_{n_4} \big] (x) \,.
\end{align}

We calculate the massless soft function to NNLO in
Appendix~\ref{sec:SoftFunction}. Our explicit calculations show that
integrals involving parton 4 vanish to this order. It seems likely to
us that this is also true at higher-orders, but do not pursue a
formal proof  here. 

\paragraph{Soft emission collinear to the top quark.}

Consider now the soft-collinear region, where $p_{X_{sc}} \sim
s_4p_3/\hat{s}$. To set up a power counting we decompose external
momenta as in the regions calculation. We can then expand the
delta-function constraint the same way:
\begin{align}
  \delta(\omega_{sc}-2p_4\cdot p_{X_{sc}}) \approx \delta(\omega_{sc}
  - \sqrt{\hat{s}}\, \bar{n}_3 \cdot p_{X_{sc}}) \, ,
\end{align}
where $\bar{n}_3 = n_4$. As for the scalar products with the gluon
field, for $i \neq 3$ we have
\begin{align}
  v_i \cdot A^a_{sc}(sv_i) \approx \frac{1}{2} \, n_i \cdot n_3 \;
  \bar{n}_3 \cdot A^a_{sc} \left( s n_i \cdot n_3 \, \frac{\bar{n}_3}{2}
  \right) .
\end{align} 
We again use the invariance of Wilson lines under the scaling $n_i\to
\lambda n_i$. With an appropriate choice of $\lambda$, the scalar
product becomes
\begin{align}
  v_i \cdot A^a_{sc}(sv_i) \to \bar{n}_3 \cdot A^a_{sc}(s\bar{n}_3)
\end{align}
\textit{irrespective} of whether $i=1,2,4$. It follows that we can
replace the product of Wilson lines as
\begin{align}
  \label{eq:Sprod}
  [\bm{Y}_{v_1} \bm{Y}_{v_2} \bm{Y}_{v_4}](x) &= \mathcal{P} \exp
  \left( ig_s \int_{0}^{\infty} ds \; \bar{n}_3 \cdot
    A_{sc}^a(x+s\bar{n}_3) \; \sum_{i\neq 3} \bm{T}_i^a \right)
  \nonumber
  \\
  &= \overline{\mathcal{P}} \exp \left( ig_s \int_{0}^{\infty} ds \;
    \bar{n}_3 \cdot A_{sc}^a(x+s\bar{n}_3) \; (-\bm{T}_3^a) \right) =
  \bm{Y}_{\bar{n}_3}^\dagger(x) \, ,
\end{align}
where we have used color conservation.
To understand the appearance of anti-path ordering in the
second line, we note that color conservation $\sum_{i\neq 3}\bm{T}_i^a
= -\bm{T}_3^a$ only applies when acting on the color singlet amplitude
directly, as in (\ref{eq:colorcons}), so one must replace, e.g.,
\begin{align}
  \left( \sum_{i\neq 3}\bm{T}_i^b \right) \left( \sum_{i\neq
      3}\bm{T}_i^a \right) = \left( \sum_{i\neq 3}\bm{T}_i^b \right)
  (-\bm{T}_3^a) = (-\bm{T}_3^a) \left( \sum_{i\neq 3}\bm{T}_i^b
  \right) = (-\bm{T}_3^a) (-\bm{T}_3^b) \, ,
\end{align}
where we have used that $\bm{T}_i$ and $\bm{T}_j$ commute when $i \neq
j$. On the other hand, when $i=3$ we just have a standard
soft-collinear Wilson line for particle 3, and no further expansion is
possible. Therefore, we can identify
\begin{align}
  \label{eq:Osc}
  \bm{O}_{sc}(x) = \bm{Y}_{v_3}(x) \, \bm{Y}^\dagger_{\bar{n}_3}(x) \, .
\end{align}
The squared matrix element involving soft collinear structure is then
\begin{align}
  \label{eq:SDdef}
  \bm{S}_D\left(\omega_{sc}, \frac{\omega_{sc}m_t}{\hat{s}},\mu\right) = \sum_{X_{sc}}
  \braket{0|\bm{O}_{sc}^\dagger(0)|X_{sc}}
  \braket{X_{sc}|\bm{O}_{sc}(0)|0} \, \delta(\omega_{sc} - \sqrt{\hat{s}}\,
  \bar{n}_3 \cdot p_{X_{sc}}) \, .
\end{align}
The functional form follows from $\bar{n}_3 \cdot v_3 = \sqrt{\hat{s}}/m_t$ along with 
properties under rescaling, similarly to (\ref{eq:scalingint}).

Since the operator $\bm{O}_{sc}$ only involves the color generator of
parton 3, the right-hand side of the above equation is diagonal in
color space. The function $\bm{S}_D$ is then proportional to the unit
matrix, namely $\bm{S}_D \equiv \bm{1} \times S_D$. To make a
connection with the literature, we use  (\ref{eq:Sconvert}) to write 
\begin{align}
  \label{eq:SDdef2}
  S_D\left(\omega_{sc}, \frac{\omega_{sc}m_t}{\hat{s}},\mu\right) = \sum_{X_{sc}}
  \braket{0|O_{sc}^\dagger(0)|X_{sc}}
  \braket{X_{sc}|O_{sc}(0)|0} \, \delta(\omega_{sc} - \sqrt{\hat{s}} \,
  \bar{n}_3 \cdot p_{X_{sc}}) \, ,
\end{align}
where
\begin{align}
  O_{sc}(x) = Y^\dagger_{v_3}(x) \, Y_{\bar{n}_3}(x) \,,
\end{align}
with the Wilson line $Y_{n}$ defined as in~(\ref{eq:Sconvert}) but with $sc$ fields only.
    
We now use the Fourier representation of the delta function to write
\begin{align}
  \label{eq:SDdef-start}
  S_D\left(\omega_{sc}, \frac{\omega_{sc}m_t}{\hat{s}},\mu\right) = \frac{1}{\sqrt{\hat{s}}} \sum_{X_{sc}} \int
  \frac{d t}{2 \pi} \, e^{i \left( \omega_{sc}/\sqrt{\hat{s}} - \bar{n}_3
      \cdot p_{X_{sc}} \right) t} \braket{0|O_{sc}^\dagger(0)|X_{sc}}
  \braket{X_{sc}|O_{sc}(0)|0} \, ,
\end{align}
and we shift the argument of
$O^\dagger_{sc}$:
\begin{align}
  \label{eq:translation}
  O^\dagger_{sc}(0)= e^{-i t \bar{n}_3 \cdot \hat{P}}
  O^\dagger_{sc}(t\bar{n}_3) \, e^{i t \bar{n}_3 \cdot \hat{P}} \, ,
\end{align}
where $\hat{P}$ is an operator acting on the external states to pick
up their momenta. This  operator 
produces a term that cancels the $p_{X_{sc}}$ dependence in the Fourier
exponent, allowing us to perform the sum over states to find 
\begin{align}
  \label{eq:SDdef-cntd}
  S_D\left(\omega_{sc}, \frac{\omega_{sc}m_t}{\hat{s}},\mu\right) &=
  \frac{1}{\sqrt{\hat{s}}} \int \frac{d t}{2 \pi} \, e^{i \omega_{sc}
    t / \sqrt{\hat{s}}}
  \braket{0|\overline{T}[O_{sc}^\dagger(t\bar{n}_3)] \,
    T[O_{sc}(0)]|0} \, ,
\end{align}
which can be shown to coincide with the soft part of the heavy-quark fragmentation
function in Eq.~(50) of Ref.~\cite{Jain:2008gb} after appropriate
replacements. In the above formula, the time-ordering $T$ and
anti-time-ordering $\overline{T}$ are imposed to guarantee the correct
ordering of the fields.\footnote{See, e.g., Appendix C of
  \cite{Becher:2007ty}.}

\paragraph{Soft emission collinear to the anti-top quark.}

Finally, we consider the $sc'$ region, where $p_{X_{sc'}} \sim
p_4s_4/m_t^2$. For the scalar products $v_i \cdot A^a_{sc'}$, $i\neq 4$,
we can perform exactly the same arguments as for the $sc$ region. The
$sc'$ region then involves the operator
\begin{align}
  O_{sc'}(x) = Y'^\dagger_{\bar{n}_4}(x) \, Y'_{v_4}(x) \, ,
\end{align}
where $\bar{n}_4=n_3$ and the Wilson line $Y'_{n}$ is defined as in~(\ref{eq:Sconvert}) but with $sc'$ fields only. As for the the delta-function constraint, there
is no possible expansion. Therefore, the matrix element for the $sc'$
region is
\begin{align}
\label{eq:Bdef}
  S_B\left(\omega_{sc'}, \frac{\omega_{sc'}}{m_t},\mu\right) =\sum_{X_{sc'}}
  \braket{0|O_{sc'}^\dagger(0)|X_{sc'}} \braket{X_{sc'}|O_{sc'}(0)|0}
  \, \delta(\omega_{sc'} - 2 m_t v_4 \cdot p_{X_{sc'}})
\end{align}
The difference between $S_B$ and $S_D$ is the time-like vector in the
delta-function constraint, as opposed to a light-like one. We can now go through the steps discussed above 
for $S_D$ to arrive at the result
\begin{align}
  \label{eq:SBdef-cntd}
  S_B\left(\omega_{sc'}, \frac{\omega_{sc'}}{m_t},\mu\right) &=
  \frac{1}{2 m_t} \int \frac{d t}{2 \pi} \, e^{i \omega_{sc'} t / (2
    m_t)} \braket{0| \overline{T}[O_{sc'}^\dagger(t v_4)] \,
    T[O_{sc'}(0)] |0} \, ,
\end{align}
which is consistent with the definition of the heavy-quark jet
function in Eq.~(46) of Ref.~\cite{Jain:2008gb}, after making the
adaptions necessary to describe a final-state antiquark.

\paragraph{ Comments.}

After inserting the matrix elements  (\ref{eq:Sdef}),
(\ref{eq:SDdef-cntd}), and (\ref{eq:SBdef-cntd}) into
(\ref{eq:Smfact1}) we arrive at the factorization formula
(\ref{eq:softfact}) for the massive soft function in the small-mass
limit.  We achieved this by studying the factorization of the
Wilson-line definition of the massive soft function in this limit.  

Another option would have been to analyze the differential cross
section in soft-collinear effective theory through a multistep
matching procedure, similarly to the analysis in
\cite{Fleming:2007qr}, where energetic top-pair production in $e^+
e^-$ collisions was studied.  In that case, after integrating out
virtualities of order $\hat{s}$ and $m_t$, one is left with two copies
of boosted HQET, which interact only through ``soft cross talk''.  In
our analysis, the $sc$- and $sc'$-momenta play the role of the
residual momenta for the two copies of boosted HQET, and the
soft momenta the role of the soft cross talk.  It is then evident that
many steps of an effective-theory analysis could be carried over from
\cite{Fleming:2007qr} and lead to the same final result.  We refer the
interested reader to that work for the set-up that could be used in such
an effective-theory analysis.

\section{Fixed-order expansions and resummation}
\label{sec:resum}
The factorization formalism derived in this work can be used in 
different ways.  The first is to view it as a tool for reformulating the 
calculation of complicated, multiscale higher-order corrections 
to the coefficient  functions $C_{ij}$ in terms of much simpler one-scale
calculations, up to corrections to  the soft and small-mass limit.
In that case, we need only fixed-order expansions of the component 
functions appearing in the factorization formula.  However, in the limit
where the  mass scales characterizing the component functions are
widely separated, for any choice of a common factorization scale $\mu_f$
the fixed-order expansion of $C_{ij}$ contains large logarithms of scale
ratios which can be resummed by deriving and solving RG equations for 
the component functions.  In this section we collect results for 
the fixed-order expansions of the component functions to NNLO, and then
discuss the structure of their RG equations.  

It is simplest to discuss higher-order corrections and RG equations in
Laplace space, where the distribution-valued functions related to 
soft real emission become simple functions, and convolutions 
reduce to multiplication. We define Laplace transforms of the component
functions as 
 \begin{align}
\label{eq:LapDefs}
\bm{\tilde{s}}_{ij}\left(\ln\frac{\hat{s}}{\bar{N}^2 \mu^2} ,x_t, \mu\right) &= \hat{s} \int_0^\infty d \lambda\,  e^{- \lambda N}  \bm{S}_{ij}\left(\lambda \hat{s}, \lambda \sqrt{\hat{s}} ,x_t, \mu\right) \, , \nonumber \\
\tilde{s}_D \left(\ln\frac{m_t}{\bar{N} \mu} ,\mu \right) &= \hat{s} \int_0^\infty d \lambda\,  e^{- \lambda N} S_D \left(\lambda \hat{s},  \lambda m_t,\mu \right) \, ,\nonumber \\
\tilde{s}_B \left(\ln \frac{\hat{s}}{\bar{N} m_t \mu} ,\mu \right) &=\hat{s} \int_0^\infty d \lambda\,  e^{- \lambda N} S_B \left(\lambda \hat{s}, \lambda \frac{\hat{s}}{m_t} ,\mu \right) \, ,
\end{align}
with $\bar{N} = N e^{\gamma_E}$.  
We can then write the Laplace-transformed hard-scattering kernels  
\begin{align}
\label{eq:LapTransf}
  \tilde{c}_{ij}(N,\hat{s},\hat{t}_1,\hat{u}_1,m_t,\mu) & 
= \hat{s} \int_0^\infty d \xi e^{- \xi N}  C_{ij}(s_4,\hat{s},\hat{t}_1,\hat{u}_1,m_t,\mu) \, ,
\end{align}
where $\xi = s_4 / \hat{s}$,\footnote{The variable $\xi$ takes values in the interval $[0,1-2 m_t/ \sqrt{\hat{s}}]$; for values of $\xi$ which are outside that interval, the integrand in (\ref{eq:LapTransf}) is considered to be zero.} 
as 
\begin{align}
\label{eq:Laplfac}
\tilde{c}_{ij}(N,\hat{s},\hat{t}_1,\hat{u}_1,m_t,\mu) & 
=    C_D^2\left(\ln\frac{m_t^2}{\mu^2},\mu \right) \Tr 
\left[\bm{H}_{ij}\left(\ln\frac{\hat{s}}{\mu^2},x_t, \mu\right) \, \bm{\tilde{s}}_{ij}\left(\ln\frac{\hat{s}}{\bar{N}^2 \mu^2} ,x_t, \mu\right) \right] \nonumber \\
&\times \tilde{s}_D \left(\ln\frac{m_t}{\bar{N} \mu} ,\mu  \right) 
\tilde{s}_B \left(\ln \frac{\hat{s}}{\bar{N} m_t \mu} , \mu \right) \nonumber \\
& + \mathcal{O}\left(\frac{\hat{s}}{N m_t^2}\right) + \mathcal{O}\left(\frac{m_t^2}{\hat{s}}\right) \,.
\end{align}

We now discuss the NNLO corrections to the various component functions above.  
The channel-independent functions $\tilde{s}_B$, $\tilde{s}_D$, and 
$C_D$, all  related to (soft) collinear emissions, are particularly simple. 
For these, we can define coefficients as  
\begin{align}
 \tilde{s}_D(L,\mu) 
&= 1 + \left( \frac{\alpha_s}{4\pi} \right) \tilde{s}_D^{(1)}(L) + \left( \frac{\alpha_s}{4\pi}\right)^2 \tilde{s}_D^{(2)}(L) 
+ {\mathcal O}(\alpha_s^3) \, ,
\label{eq:ColExp}
\end{align}
and similarly for $\tilde{s}_B$ and $C_D$.  Compact results for all
of these functions to NNLO can be extracted from the literature, and
are gathered in Appendix~\ref{sec:coeffs}.

For the channel-dependent, matrix-valued massless hard and soft functions,
we define perturbative expansion coefficients
\begin{align}
  \bm{H} &= \alpha_s^2 \, \frac{3}{8d_R} \left[ \bm{H}^{(0)} + \left( \frac{\alpha_s}{4\pi} \right) \bm{H}^{(1)} + \left( \frac{\alpha_s}{4\pi} \right)^2 \bm{H}^{(2)}  + {\mathcal O}(\alpha_s^3) \right] , \nonumber
  \\
  \tilde{\bm{s}} &= \tilde{\bm{s}}^{(0)} + \left( \frac{\alpha_s}{4\pi} \right) \tilde{\bm{s}}^{(1)} + \left( \frac{\alpha_s}{4\pi} \right)^2 \tilde{\bm{s}}^{(2)} + {\mathcal O}(\alpha_s^3) \, . \label{eq:HSexp}
\end{align}
Here and in the remainder of the section we suppress the subscript
indicating the channel dependence of the hard and soft functions, as
well as their explicit arguments.  While the NNLO hard functions
$\bm{H}^{(2)}$ are unknown, the quantities ${\rm
  Tr}\left[\bm{H}^{(2)}\tilde{\bm{s}}^{(0)} \right]$ were recently
extracted in \cite{Ferroglia:2013zwa}, using NNLO corrections from
massless two-to-two scattering obtained in \cite{Anastasiou:2000kg,
  Anastasiou:2000mv, Anastasiou:2001sv, Glover:2001af, Glover:2001rd}
along with a subtraction procedure.  The rather lengthy expressions
can be found in electronic form with the arXiv version of that paper.
The massless soft functions are not available in the literature, but
we construct results up to NNLO in
Appendix~\ref{sec:SoftFunction}. Here again the results are lengthy,
and are included in {\sc mathematica} files with the arXiv submission
of the present paper.

We can make use of these results to form approximations to the
Laplace-transformations of the expansion coefficients of the
hard-scattering kernels defined in (\ref{eq:Cexp}).  To leading order
in the soft and small mass limits, we have
\begin{align}
  \tilde{c}^{(0)} &= \frac{3}{8 d_R} \Tr \left[ \bm{H}^{(0)} \tilde{\bm{s}}^{(0)}  \right] , \nonumber
  \\
  \tilde{c}^{(1)} &= \frac{3}{8 d_R} \Tr \left[ \bm{H}^{(1)} \tilde{\bm{s}}^{(0)} + \bm{H}^{(0)} \tilde{\bm{s}}^{(1)} + \left(\tilde{s}_B^{(1)} +  \tilde{s}_D^{(1)}+ 2 C_D^{(1)} \right) \bm{H}^{(0)} \tilde{\bm{s}}^{(0)}  \right] , \nonumber
  \\
  \tilde{c}^{(2)} &=\frac{3}{8 d_R} \Tr \Biggl\{ \bm{H}^{(2)} \tilde{\bm{s}}^{(0)} + \bm{H}^{(0)} \tilde{\bm{s}}^{(2)} + \bm{H}^{(1)} \tilde{\bm{s}}^{(1)} + \left( \tilde{s}_B^{(1)}+  \tilde{s}_D^{(1)}+ 2 C_D^{(1)}  \right) \left( \bm{H}^{(0)} \tilde{\bm{s}}^{(1)} + \bm{H}^{(1)} \tilde{\bm{s}}^{(0)} \right) \nonumber
  \\
  &+ \bigg[ \tilde{s}_B^{(2)} +  \tilde{s}_D^{(2)} + 2 C_D^{(2)} + \left(C_D^{(1)}\right)^2 + \tilde{s}_B^{(1)} \tilde{s}_D^{(1)}   
+ 2 C_D^{(1)}
  \left(\tilde{s}_B^{(1)}+ \tilde{s}_D^{(1)} \right)\bigg] \bm{H}^{(0)} \tilde{\bm{s}}^{(0)} \Biggr\} \, .
  \label{eq:cts}
\end{align}
The above result for the soft and small-mass limit of the 
NNLO coefficient $\tilde{c}^{(2)}$ is  particularly interesting 
because the exact coefficient in fixed-order perturbation theory is unknown.  
Approximations to this coefficient based on soft -gluon resummation to 
NNLL order for arbitrary $m_t$ were derived in
\cite{Kidonakis:2010dk, Ahrens:2011mw}.  To explain how the results given
here go beyond those works, we define an explicit expansion of the 
coefficient function as
 \begin{align}
  \tilde{c}^{(2)}(N, \hat{s},\hat{t}_1,\hat{u}_1,m_t ,\mu_f) &= \sum_{n=0}^4
  c^{(2,n)}(\hat{s},\hat{t}_1,\hat{u}_1,m_t ,\mu_f) \, \ln^n\frac{\hat{s}}{\bar{N}^2\mu_f^2} +
  \mathcal{O}\left(\frac{1}{N}\right) .
\end{align}
The NNLO approximations from \cite{Kidonakis:2010dk, Ahrens:2011mw}
determine the coefficients $c^{(2,n)}$ with $n=1\dots 4$, as exact functions
$m_t$.  From the viewpoint of a fixed-order expansion, the
results for these coefficients in the small-mass limit, determined
from (\ref{eq:cts}), do not offer an improvement.  However, it is a
very non-trivial check on the factorization formalism that the 
coefficients derived above agree with the small-mass limit of those
from \cite{Kidonakis:2010dk, Ahrens:2011mw}, a fact which we have
confirmed.  

The NNLO approximations from \cite{Kidonakis:2010dk, Ahrens:2011mw} do
not determine $c^{(2,0)}$, as this coefficient is formally of NNNLL
order.  The expansion (\ref{eq:cts}) determines it in the small-mass
limit (up to corrections involving heavy-quark loops, which we return
to in the next section). It will thus be interesting to study the
numerical implications of our results for high-$p_T$ top production,
where corrections to the small-mass limit are negligible and the extra
terms calculated here offer a clear improvement on the NNLO
approximations from \cite{Kidonakis:2010dk, Ahrens:2011mw}.  In fact,
our results form the basis for a full NNLO  soft plus virtual
approximation in the small mass limit, meaning that they determine also the
delta-function coefficient in (\ref{eq:C2}).

The convergence of the fixed-order expansion discussed above can be
invalidated when the logarithms of scale ratios are large.  In that
case, one must resum the logarithms by deriving and solving RG
equations.  The RG equations for the channel-independent functions 
read
\begin{align}
\frac{d}{d\ln\mu} \tilde{s}_B\left(\ln\frac{m_t}{\mu},\mu \right)
& = \left(-C_F \gamma_{\rm cusp}(\alpha_s)\ln\frac{m_t^2}{\mu^2} 
+ 2\gamma^B(\alpha_s) \right) \tilde{s}_B\left(\ln\frac{m_t}{\mu},\mu\right) \, , \label{eq:RGEsB}\\
\frac{d}{ d \ln \mu} \tilde{s}_D \left(\ln\frac{m_t}{\mu}, \mu\right) & =  \left(C_F \gamma_{\rm cusp}(\alpha_s)\ln\frac{m_t^2}{\mu^2} 
-2 \gamma^S(\alpha_s) \right) \tilde{s}_D\left(\ln\frac{m_t}{\mu},\mu\right) \, , \label{eq:RGEsD}\\
\frac{d}{ d \ln \mu} C_D \left(\ln\frac{m_t^2}{\mu^2}, \mu\right)& =  \left( - C_F \gamma_{\rm cusp}(\alpha_s)\ln\frac{m_t^2}{\mu^2} 
+2 \gamma^S(\alpha_s)  + 2 \gamma^{\phi_q} \right) C_D\left(\ln\frac{m_t^2}{\mu^2},\mu\right) \, .
\label{eq:RGECD}
\end{align}  
Perturbative results for the anomalous dimensions to order
$\alpha_s^2$ using the above definitions (and that of the
$\gamma^\phi$ in (\ref{eq:APev}) below) are listed in
Appendix~\ref{sec:coeffs}.  The RG equations for the matrix valued
hard and soft functions have a similar form.  We write these as
\begin{align}
  \label{eq:Hev}
  \frac{d}{d\ln\mu} \bm{H}\left(\ln\frac{\hat{s}}{\mu^2},x_t,\mu\right)
 &= \bm{\Gamma}_H\left(\ln\frac{\hat{s}}{\mu^2},x_t,\mu\right) \,
  \bm{H}\left(\ln\frac{\hat{s}}{\mu^2},x_t,\mu\right)
 \nonumber \\
 & \hspace{0.9cm} +
\bm{H}\left(\ln \frac{\hat{s}}{\mu^2},x_t,\mu\right) \, 
  \bm{\Gamma}_H^\dagger\left(\ln \frac{\hat{s}}{\mu^2},x_t,\mu\right) 
\, , \nonumber \\
    \frac{d}{d\ln\mu} \, \tilde{\bm{s}}\left(\ln\frac{\hat{s}}{\mu^2},x_t,\mu\right) &= 
 -   \bm{\Gamma}^{\dagger}_S\left(\ln \frac{\hat{s}}{\mu^2},x_t ,\mu\right)\tilde{\bm{s}}\left(\ln\frac{\hat{s}}{\mu^2},x_t,\mu\right)
 \nonumber \\
 & \hspace{0.9cm} -
    \tilde{\bm{s}}\left(\ln\frac{\hat{s}}{\mu^2},x_t,\mu\right)\bm{\Gamma}_S\left(\ln \frac{\hat{s}}{\mu^2},x_t ,\mu\right) \, .
\end{align}
The anomalous dimension for the hard matrix is 
\begin{align}
  \label{eq:gammaH}
  \bm{\Gamma}_H\left(\ln\frac{\hat s}{\mu^2},x_t,\mu\right) = A(\alpha_s) \left( \ln\frac{\hat{s}}{\mu^2} -
    i\pi \right) \bm{1} + \bm{\gamma}^h(x_t,\alpha_s) \, ,
\end{align}
where $A=2C_F\gamma_{\text{cusp}}$ in the $q\bar q$ channel and
$A=(N+C_F)\gamma_{\rm cusp}$ in the $gg$ channel.  That for the soft
function has the form
  \begin{align}
  \label{eq:gammaS}
 \bm{\Gamma}_S\left(\ln \frac{\hat{s}}{\mu^2},x_t ,\mu\right)= A_s(\alpha_s) \left( \ln\frac{\hat{s}}{\mu^2} -
    i\pi \right) \bm{1} + \bm{\gamma}^s(x_t, \alpha_s) \, .
\end{align}
The coefficients $A_s$ and  $\bm{\gamma}^s$ can be derived from the 
results above, along with the condition that the $\mu$-dependence 
of the partonic cross section cancels against that in the PDFs to give
a $\mu$-independent hadronic cross section. We write the Altarelli-Parisi
kernels in the soft limit as
 \begin{align} \label{eq:APev}
  P_{ii}(z,\mu) = \frac{2 C_i \gamma_{\rm cusp}(\alpha_s)}{(1-z)_+}+2
  \gamma^{\phi_i}(\alpha_s)\delta(1-z) \, .
\end{align}
We then find $A_s= C_F\gamma_{\rm cusp}$ in the $q\bar q$ channel and
$A_s=C_A\gamma_{\rm cusp} $ in the $gg$ channel, and that 
\begin{align}
  \bm{\gamma}^s(x_t,\alpha_s) &= \bm{\gamma}^h(x_t,\alpha_s) +
  \big[2\gamma^{\phi}(\alpha_s) + 2\gamma^{\phi_q}(\alpha_s) 
\nonumber \\
&
\qquad \qquad +\gamma^B(\alpha_s)+\gamma^{S}(\alpha_s)
-A_s(\alpha_s) \ln x_t(1-x_t)\big]
  \bm{1} \, . \label{eq:gs}
\end{align}
The term proportional to $\ln x_t (1-x_t)$ is needed to cancel the 
$\mu$-dependence of the PDFs  and follows from the 
derivation given in Section~3.2 of \cite{Ahrens:2011mw}.

These RG equations can be solved in the standard way, and in fact many
of the ingredients can be recycled directly from
\cite{Ferroglia:2012ku} after appropriate replacements.  The
perturbative components gathered here form a starting point for an
analysis of a simultaneous small-mass and soft-gluon resummation to
NNLL.  We plan to return to a phenomenological analysis of resummation
effects, and a comparison with the fixed-order approximations defined
above, in future work.

\section{Heavy-quark loops}
\label{sec:top_loops}

A technical subtlety we now address is the treatment of closed heavy-quark
loops.  The way in which they contribute to the factorization formula
is determined by our parametric counting $s_4 \ll m_t^2$.  Since in
that case the heavy-quark mass is much larger than any of the scales
characteristic of real radiation, it is not possible for a soft gluon
to split into on-shell top quarks.  Therefore, heavy-quark loops do
not contribute to any of the functions in (\ref{eq:softfact}) related to
soft real emission.  Heavy-quark loops decouple from these functions,
and the correct prescription is to evaluate them in a theory with five
massless flavors.

For the virtual corrections, on the other hand, there is no such 
decoupling of heavy-quark loops, and one must include them 
explicitly through diagrammatic computations.  This leads to a 
modification of (\ref{eq:hardfact}). In general, we can write
\begin{align}
\label{eq:hardfactNh}
 \bm{H}^m_{ij}(\hat{s},\hat{t}_1,\hat{u}_1,m_t,\mu_f)=
 \bm{C}^{ij}_h(\hat{s},x_t,m_t,\mu_f) 
C_D^2(m_t,\mu_f)  \bm{H}_{ij}(\hat{s}, x_t, \mu_f)
+{\cal O}\left(\frac{m_t^2}{\hat{s}}\right)\,.
\end{align}  
The notation is such that the $\bm{C}^{ij}_h$ contains any explicit
dependence on $n_h=1$ and represents the effects of closed top-quark
loops, while the second and third factor in the r.h.s. of
(\ref{eq:hardfactNh}) are the same as in (\ref{eq:hardfact}) and are
evaluated with $n_l=5$ light flavors. 

A few words about the renormalization schemes are in order. The hard functions are the Wilson coefficients arising when matching from the amplitudes in full QCD to the ones in the effective theory. To obtain the finite hard functions, we need to perform several renormalizations. These include the usual renormalization of the strong coupling constant, the quark masses, the quark and gluon propagators in full QCD, as well as the renormalization of the operators in the effective theory. We renormalize the masses and the propagators in the on-shell scheme, and renormalize the effective operators in the $\msbar$ scheme. Furthermore, we renormalize the running coupling constant
in the $\msbar$ scheme with five active flavors (in practice by first performing renormalization
with six active flavors and then applying a decoupling transformation).

An interesting question is whether the matching coefficient $\bm{C}_h$
in (\ref{eq:hardfactNh}) can be factorized into one-scale functions,
whether it is diagonal in color space, and whether it depends on the
channel.  In the absence of heavy-quark loops, the $m_t$ dependence is
contained solely in $C_D$, and is related to regions of loop momenta
collinear to $p_3$ and $p_4$. Top-quark loops can introduce an $m_t$
dependence even in diagrams not involving $p_3$ or $p_4$, and can
otherwise change the regions analysis in such a way that not all $m_t$
dependence is related to collinear regions.  A full analysis of these
higher-order corrections is beyond the scope of the paper.  However,
we note that the two-loop corrections depending on $n_h$ were
calculated in the small-mass limit in \cite{Czakon:2007ej,
  Czakon:2007wk}, and do not appear to factorize in a simple way, as
pointed out in \cite{Mitov:2006xs}.  The same is true of the analogous
NNLO corrections calculated for Bhabha scattering in
\cite{Becher:2007cu}.

In any case, the closed heavy-quark loops can be included in
fixed-order perturbation theory by calculating the contributions
involving powers of $n_h$ to the massive hard function in the
small-mass limit. The NLO results can be extracted from
\cite{Ahrens:2010zv}. To obtain the NNLO results as a matrix would be
rather involved. However, we hope to extract the contribution of the
$n_h$-dependent part of NNLO hard function to the coefficient function
(\ref{eq:fac}) using the the method followed in
\cite{Ferroglia:2013zwa} in future work.  This requires two main
pieces related to $n_h$-dependent corrections.  The first is the UV
renormalized NNLO virtual corrections in the small-mass limit.  The
two-loop contributions were given in \cite{Czakon:2007ej,
  Czakon:2007wk}, but the one-loop squared pieces are not readily
available in the literature. In addition, one must determine certain
color-decomposed one-loop amplitudes to order $\epsilon^2$.  With
these building blocks in place, one can calculate the finite remainder of
the  $n_h$-dependent terms and add them 
to the results gathered in the previous section to achieve a full soft
plus virtual approximation at NNLO in the double soft and small-mass
limit.

\section{Conclusions}
\label{sec:conclusions}

We have derived a novel factorization formula appropriate for the
double soft and small-mass limits of single-particle inclusive cross
sections in top-quark pair production at hadron colliders.  This
formula applies to double differential distributions in the rapidity
and $p_T$ of the heavy top (or anti-top) quark within this limit. In
the absence of closed heavy-quark loops in virtual corrections, we found that
the partonic cross section factorizes into five component functions,
each depending on a single momentum scale.

Our method was to start from the factorization formula (\ref{eq:fac})
for top-quark pair production in the soft limit and then subfactorize
the component parts as appropriate for the small-mass limit.  For
virtual corrections contained in the so-called hard function the
method for doing this was already available in the literature.  On the
other hand, our result (\ref{eq:softfact}) for the factorization of
the soft function, related to real emission in the double soft and
small-mass limits, is new.  We motivated the result by first
performing a diagrammatic factorization of real emission using the
method of regions in Section~\ref{sec:phsp}. Our analysis revealed
that three types of soft radiation contribute: radiation
simultaneously soft and collinear to the observed top-quark, radiation
soft and collinear to the unobserved anti-top quark, and wide angle
soft emission. In Section~\ref{sec:facmasssoft}  we showed how to factorize the general Wilson-loop
operator definition of the soft function into three component
Wilson-loop operators related to these regions.
We demonstrated explicitly that the two types of collinear operators
are diagonal in color space and connected the Wilson loop operators
with the soft part of the heavy-quark fragmentation function and the
heavy-quark jet function introduced in \cite{Fleming:2007qr}. The
wide-angle soft emission goes into a ``massless soft function''
defined in (\ref{eq:Sdef}). It involves a
non-trivial matrix structure characteristic of soft emissions in
two-to-two scattering.

Most of the component functions entering our factorization formula
could be extracted to NNLO from results in the literature.  We added
to this literature by computing the NNLO massless soft function.  We
showed that the anomalous dimensions appearing in the
renormalization-group equation for the NNLO function is consistent
with the factorization formalism to this order, providing a strong
consistency check on the factorized formula as well as our
higher-order perturbative computation. An equivalent check, described
in Section~\ref{sec:resum}, is that the NNLO logarithmic corrections obtained
by expanding out the factorization formula 
agree with the small-mass limit of those determined by NNLL soft-gluon
resummation in \cite{Kidonakis:2010dk, Ahrens:2011mw}.  Our results
provide nearly all elements for the construction of an NNLO soft plus
virtual approximation to the differential cross section in the
small-mass limit. The missing piece is the NNLO virtual corrections
related to closed heavy-quark loops, which we hope to  calculate in future
work.

We expect that the results obtained here will provide useful 
insight into the structure of higher-order QCD corrections to 1PI
observables in the boosted regime. On the one hand, the NNLO 
soft plus virtual approximation can be used to study numerically to 
what extent logarithmic soft gluon corrections dominate over non-logarithmic
delta-function terms, thus assessing the reliability of the NNLO 
approximations \cite{Kidonakis:2010dk, Ahrens:2011mw} for boosted production.
On the other hand, our factorization formalism provides the starting
point for a simultaneous resummation of small-mass and soft logarithms 
in the partonic cross section to NNLL.

\section*{Acknowledgments}
This work is supported in part by the National Natural Science
Foundation of China under Grant No. 11345001. The work of A.F. was
supported in part by the PSC-CUNY Award No. 66590-00-44 and by the
National Science Foundation Grant No. PHY-1068317. S.M. is supported the
UK's STFC.  B.P. is grateful to KITP Santa Barbara for hospitality and
support, and B.P. and S.M. would like to thank the ESI Vienna for 
hospitality and support.  

\appendix

\section{The massless soft function to NNLO}
\label{sec:SoftFunction}
The calculation of the massless soft function (\ref{eq:Sdef}) proceeds
similarly to \cite{Ferroglia:2012uy}.   The   difference in the present case is that 
the constraint vector is light-like instead of time-like, but since the basic phase-space
integrals were calculated for an arbitrary constraint vector, this offers no additional complication.
In fact,   the end results are slightly 
simpler, and integrals involving parton 4 vanish.  

We first go through the NLO calculation as an example.  
We obtain the bare NLO soft function through the following sum over legs:
\begin{align}
  \bm{S}^{(1)}_{\rm bare} &= \frac{2}{\omega}
  \left(\frac{\sqrt{\hat{s}}\mu}{\omega}\right)^{2\epsilon} \, \sum_{\text{legs $\neq$ 4}} \,
  \bm{w}_{ij}^{(1)} \, \bar{I}_1(a_{ij}) \,\nonumber
  \\
  &= \frac{4}{\omega} \left(\frac{\sqrt{\hat{s}}\mu}{\omega}\right)^{2\epsilon}
  \left( \bm{w}_{12}^{(1)} \, \bar{I}_1(a_{12})+ \bm{w}_{13}^{(1)} \, \bar{I}_1(a_{13}) +  \bm{w}_{23}^{(1)} \, \bar{I}_1(a_{23})\right)\, .
\end{align}
Explicit results for the matrices $\bm{w}_{ij}^{(1)}$ can be found in
\cite{Ferroglia:2012uy}.  The integral $I_1$ is defined as\footnote{This integral is identical to the one found in Eqs.~(13-14) of  \cite{Becher:2012za}.}
\begin{align}
I_{1}(a_{ij}) &= \int d^dk \ \delta^{+}(k^2) \delta(\omega - \sqrt{\hat{s}} n_4 \cdot k ) \ \frac{n_i \cdot n_j}{n_i \cdot k \, n_j \cdot k} \, , \label{eq:I1}
\end{align}
while the stripped integral $\bar{I}_1$ can be obtained from (\ref{eq:I1}) through the relation
\begin{align}
I_{1}(a_{ij}) &= \pi^{1-\epsilon} e^{-\epsilon \gamma_{\mbox{\tiny E}}} \omega^{-1-2 \epsilon} \hat{s}^{\epsilon} \bar{I}_1(a_{ij}) \, .
\end{align}
The general result for the stripped integral
resulting from gluon emissions associated with partons $ij$ is
\begin{align}
  \bar{I}_1(a_{ij}) &= \frac{2 \, e^{\epsilon\gamma_E} \,
    \Gamma(-\epsilon)}{\Gamma(1-2\epsilon)} \, a_{ij}^{-\epsilon} \;
  {}_2F_1(-\epsilon,-\epsilon,1-\epsilon,1)  = -2 e^{\epsilon\gamma_E} a_{ij}^{-\epsilon}  \frac{\Gamma(\epsilon) \Gamma^2(1-\epsilon)}{\Gamma(1-2 \epsilon)}\,,\label{eq:I1res}
\end{align}
where
\begin{align}
a_{ij}=\frac{2n_i\cdot n_j}{n_4\cdot n_i n_4 \cdot n_j}.
\end{align}
 (Note that $a_{ij}$ is different from the one defined in \cite{Ferroglia:2012uy}.)
With this definition, we have 
\begin{align}
a_{12} & =\frac{1}{x_t(1-x_t)}, \qquad
a_{13}  =\frac{x_t}{1-x_t}, \qquad
a_{23}  =\frac{1-x_t}{x_t}.
\end{align}
The stripped integral can easily be expanded in $\epsilon$ (and shown 
to be equivalent to $I_{12}$ from \cite{Ahrens:2011mw}, which contrary to first
appearance does not depend on $m_t$ when Laplace transformed with respect to 
$s_4/\sqrt{\hat{s}}$ instead of $s_4/m_t$). We observe that the $\epsilon$-expansion of the integral $I_1(a_{ij})$ involves only
logarithms of the argument, since the ${}_2F_1$ function does not depend on $a_{ij}$:
\begin{align}
\bar{I}_1(a_{ij}) &= -\frac{2}{\epsilon} + 2 \ln (a_{ij}) + \left( \frac{\pi^2}{6} - \ln^2{(a_{ij})} \right) \epsilon +
\frac{1}{6} \left( 28 \zeta(3) + 2 \ln^3 (a_{ij}) - \pi^2 \ln(a_{ij})  \right) \epsilon^2 + {\mathcal O}(\epsilon^3) \, .
\end{align}

The NNLO contributions read (taking account that attachments to parton 4 vanish)
\begin{align}
  \label{eq:S2bare}
  \bm{S}^{(2)}_{\text{bare}} &= 
\frac{4}{\omega} \left(\frac{\sqrt{\hat{s}}\mu}{\omega} \right)^{4\epsilon}
  \bigg[2 \bm{w}^{(1)}_{12} \left( \bar{I}_2(a_{12}) + C_A\bar{I}_6(a_{12}) + C_A
      \bar{I}_{7,1}(a_{12}) + C_A \bar{I}_{7,2}(a_{12}) \right)
   \\
  &+ 2\bm{w}^{(1)}_{13} \left( \bar{I}_2(a_{13}) + C_A
    \bar{I}_6(a_{13}) + C_A \bar{I}_{7,1}(a_{13}) + C_A
    \bar{I}_{7,2}(a_{13}) \right) \nonumber \\
  &+ 2\bm{w}^{(1)}_{23} \left(\bar{I}_2(a_{23}) + C_A
    \bar{I}_6(a_{23}) + C_A \bar{I}_{7,1}(a_{23}) + C_A
    \bar{I}_{7,2}(a_{23}) \right) \nonumber \\
  &+ 2 (\bm{w}^{(3)}_{12} \bar{I}_3(a_{12}) +\bm{w}^{(3)}_{13} \bar{I}_3(a_{13})
 +\bm{w}^{(3)}_{23} \bar{I}_3(a_{23})) 
\nonumber \\
& + \bm{w}^{(4)}_{12} \left( \bar{I}_4(a_{12}) + 2\bar{I}_5(a_{12})\right)
+ \bm{w}^{(4)}_{13} \left(\bar{I}_4(a_{13}) + 2\bar{I}_5(a_{13})\right)
+ \bm{w}^{(4)}_{23} \left(\bar{I}_4(a_{23}) + 2\bar{I}_5(a_{23})\right) 
\nonumber \\
& + 2\bm{w}^{(8)}_{123}\bar{I}_8(a_{12},a_{13})
+2\bm{w}^{(8)}_{213}\bar{I}_8(a_{12},a_{23})+
2\bm{w}^{(8)}_{312}\bar{I}_8(a_{13},a_{23})\bigg]
 \, . \nonumber
\end{align}
The integrals $\bar{I}_i$ ($i \in \{2,\cdots,8\}$) are the same as the corresponding integrals in \cite{Ferroglia:2012uy}, except  that with respect to the results in that paper one should replace the integral argument $a$ according $a \to 1-a$ in order to fit the notation of the present work, and then expand to leading order for $a \to 0$.

The bare function has poles in $\epsilon$ which must be removed through a renormalization procedure.
This is most easily done in Laplace space.  It is straightforward to obtain the bare Laplace-transformed
function  by performing the integral in the definition (\ref{eq:LapDefs}).  The form
of the RG equation (\ref{eq:Hev}) implies that the renormalized function can be found through the relation
\begin{align}
 \tilde{\bm{s}}= \bm{Z}_s^{\dagger}\tilde{\bm{s}}_{\rm bare}\bm{Z}_s  \, ,
   \label{eq:ren} 
\end{align}
where the renormalization factor $\bm{Z}_s$ reads 
\begin{align}
\ln \bm{Z}_s &= \frac{\alpha_s}{4 \pi} \left(- \frac{A_{s,0}}{2 \epsilon^2} + \frac{A_{s,0} L + \bm{\gamma}^s_0}{2 \epsilon} \right) \nonumber \\
& + \left( \frac{\alpha_s}{4 \pi} \right)^2 \left[ \frac{3 A_{s,0} \beta_0}{8 \epsilon^3} - \frac{A_{s,1} 
+2 \beta_0 \left( A_{s,0} L +  \bm{\gamma}^s_0 \right)}{8 \epsilon^2}  + 
\frac{A_{s,1} L +  \bm{\gamma}^s_1}{4 \epsilon}\right] + \cdots .
\end{align}
We have defined expansion coefficients 
\begin{align}
A_s(\alpha_s) = \frac{\alpha_s}{4 \pi} A_{s,0} + \left( \frac{\alpha_s}{4 \pi} \right)^2 A_{s,1} + \cdots\, ,
\label{eq:expAs}
\end{align}
and similarly for the other anomalous dimensions.  Note that  $\bm{\gamma}^s$ depends on the
$\bm{\gamma}^h$ through (\ref{eq:gs}).  An explicit expression for $\bm{\gamma}^h (x_t,\alpha_s)$  can 
be found in Appendix~\ref{sec:coeffs}.

By evaluating (\ref{eq:ren}) using the above expressions, and by renormalizing the bare coupling constant which appears in $\tilde{\bm{s}}_{\rm bare}$
in the $\overline{\text{MS}}$ scheme
 by replacing
\begin{align}
\alpha_s^{{\tiny \mbox{bare}}} = e^{\epsilon\gamma_E}(4\pi)^{-\epsilon}\mu^{2\epsilon}
\alpha_s\left(1 -\frac{\alpha_s}{4\pi} \frac{\beta_0}{\epsilon}  + {\mathcal O} (\alpha_s^2)\right) \, ,
\end{align}
one finds that
the renormalized soft function on the left-hand side is indeed
finite. 
 This provides a
strong check on the validity of the factorization formula used to
derive the RG equation, and also on our calculation of the bare NNLO
function.
 Specifically, by expanding everything in powers of $\alpha_s/4 \pi$, at NLO and NNLO one finds
\begin{align}
\tilde{\bm{s}}^{(1)}(L,x_t) &= \tilde{\bm{s}}^{(1)}_{\rm bare} + \bm{Z}^{\dagger (1)}_s \tilde{\bm{s}}^{(0)}+ 
\tilde{\bm{s}}^{(0)} \bm{Z}^{ (1)}_s \, ,\nonumber \\
\tilde{\bm{s}}^{(2)}(L,x_t) &= \tilde{\bm{s}}^{(2)}_{\rm bare} + \bm{Z}^{\dagger (2)}_s \tilde{\bm{s}}^{(0)} + 
\tilde{\bm{s}}^{(0)} \bm{Z}^{ (2)}_s + \bm{Z}^{\dagger (1)}_s \tilde{\bm{s}}^{(1)}_{\rm bare} + 
\tilde{\bm{s}}^{(1)}_{\rm bare} \bm{Z}^{ (1)}_s+\bm{Z}^{\dagger (1)}_s \tilde{\bm{s}}^{(0)} \bm{Z}^{(1)}_s -
\frac{\beta_0}{\epsilon} \tilde{\bm{s}}^{(1)}_{\rm bare} \, .
\end{align}

The renormalized soft functions can be written in terms of logarithms of arguments $x_t$ and $1- x_t$. We list here the results for the renormalized NLO soft function in the $q \bar{q}$ and $gg$ production channels. The specific form 
of the soft matrix depends on the choice of the color basis.
In this paper we employ the $s$-channel singlet-octet basis (see for example (31) in \cite{Ferroglia:2013zwa}). 
For the sake of brevity we set $N_c = 3$ and take into account that the soft function is symmetric. In the following, we indicate the element $ij$ of the matrix $\tilde{\bm{s}}^{(n)}_{k}$ ($k \in \{q\bar{q}, g g\}$) as $\tilde{s}^{(n)}_{k,ij}$. In the quark annihilation channel one finds
\begin{align}
\tilde{s}^{(1)}_{q \bar{q},11} &= 8 L^2-16 L \left(\ln (1-x_t)+\ln (x_t)\right)+8 \left( \ln(1-x_t)+\ln (x_t) \right)^2+4 \pi ^2 \, ,\nonumber \\
\tilde{s}^{(1)}_{q \bar{q},12} &= \frac{16}{3} L \left(\ln (x_t)- \ln (1-x_t) \right)\, ,\nonumber \\
\tilde{s}^{(1)}_{q \bar{q},22} &= \frac{16 L^2}{9}-\frac{8}{9} L \left(2  \ln (1-x_t)-\ln (x_t) \right)+\frac{16}{9} \ln ^2(1-x_t)+\frac{16 \ln^2(x_t)}{9}\nonumber \\
&-\frac{40}{9} \ln (1-x_t) \ln(x_t)+\frac{8 \pi ^2}{9} \, .
\end{align}
For the gluon fusion channel one finds 
\begin{align}
\tilde{s}^{(1)}_{gg,11} &= 18 L^2-36 L\left( \ln (1-x_t)+\ln (x_t)\right)+18 \left( \ln(1-x_t) + \ln(x_t)\right)^2 +9 \pi ^2\, ,\nonumber \\
\tilde{s}^{(1)}_{gg,12} &=12 L \left( \ln (x_t)-\ln (1-x_t)\right) \, ,\nonumber \\
\tilde{s}^{(1)}_{gg,13} &= 0\, ,\nonumber \\
\tilde{s}^{(1)}_{gg,22} &=9 L^2-9 L \left( \ln (1-x_t)+ \ln
   (x_t) \right) +9 \ln ^2(1-x_t)+9 \ln
   ^2(x_t)+\frac{9 \pi ^2}{2} \, ,\nonumber \\
\tilde{s}^{(1)}_{gg,23} &=5 L \left( \ln (x_t)- \ln (1-x_t) \right) \, ,\nonumber \\
\tilde{s}^{(1)}_{gg,33} &=  5 L^2-5 L\left( \ln (1-x_t)+ \ln (x_t) \right)+5 \ln
   ^2(1-x_t)+5 \ln ^2(x_t)+\frac{5 \pi ^2}{2} \, .
\end{align}

The matrix elements of the NNLO soft matrices have longer analytic expressions. The interested reader can find them in the form of {\sc mathematica} input files included in the arXiv submission of the present work.

\section{Matching coefficients and anomalous dimensions}
\label{sec:coeffs}

Here we list the matching coefficients and anomalous dimensions 
appearing in Section~\ref{sec:resum}.  

We first list results for the coefficients in (\ref{eq:ColExp}) as well the corresponding coefficients in the expansions
of $\tilde{s}_B$ and $C_D$.  Using \cite{Becher:2005pd}, we find
(for simplicity, here and below we set the number of colors to $N_c=3$  in the NNLO coefficient) 
\begin{align}
  \tilde{s}_D^{(1)}(L/2) &= -C_F \left( L^2+2L+
  \frac{5\pi^2}{6} \right)\, ,
  \\
  \tilde{s}_D^{(2)}(L/2) &= \frac{8}{9}L^4 +
  \left(\frac{76}{9}-\frac{8}{27}n_l\right)L^3 +
  \left(-\frac{104}{9}+\frac{76\pi^2}{27}+\frac{16}{27} n_l\right)L^2
  \nonumber
  \\
  & + \left(\frac{440}{27}+\frac{416\pi^2}{27}- 72\zeta_3
    +\frac{16}{81}n_l -\frac{16\pi^2}{27} n_l\right)L \nonumber
  \\
  &-\frac{1304}{81}-\frac{233\pi^2}{9} +\frac{1213\pi^4}{405}
  -\frac{1132\zeta_3}{9}+\left(-\frac{16}{243}+\frac{14\pi^2}{27}
    +\frac{88\zeta_3}{27}\right)n_l  \, ,
\end{align}
where $n_l$ indicates the number of active flavors.
The matching coefficient $C_D$ can be written as 
 \begin{align}
  C_D^{(1)}(L) & = C_F\left(L^2 - L +
    4+\frac{\pi^2}{6}\right)\,,
  \\
  C_D^{(2)}(L) & = \frac{8}{9}L^4 -
  \left(\frac{20}{3}-\frac{8}{27}n_l\right)L^3 +
  \left(\frac{406}{9}-\frac{28\pi^2}{27}-\frac{52}{27} n_l\right)L^2
  \nonumber
  \\
  & - \left(\frac{2594}{27}+\frac{248\pi^2}{27}- \frac{232\zeta_3}{3}
    -\frac{308}{81}n_l -\frac{16\pi^2}{27} n_l\right)L \nonumber
  \\
  &+\frac{21553}{162}+\frac{107\pi^2}{3}-\frac{749\pi^4}{405}
  +\frac{260\zeta_3}{9} +\frac{16\pi^2}{9} \ln 2
  -\left(\frac{1541}{243}+\frac{74\pi^2}{81}
    +\frac{104\zeta_3}{27}\right)n_l   \nonumber \\
& -\delta_C  4\pi^2 C_A C_F \, .
\end{align}
The NNLO coefficient was originally extracted in \cite{Neubert:2007je}
using the result for $\tilde{s}_D$ along with the 
NNLO result for the heavy-quark fragmentation function calculated in
\cite{Melnikov:2004bm}, and yields the above equation with $\delta_C=0$
in the last line.  It was extracted directly using the relationship between
small-mass and massless amplitudes
in the Appendix of \cite{Ferroglia:2012ku}, which yields the above result
with  $\delta_C=1$.  The discrepancy between the two methods of extracting
the NNLO coefficient remains unresolved.

Finally,  the Laplace transformed heavy-quark jet function is easily derived from results given in \cite{Jain:2008gb}.  Explicitly, 
 \begin{align}
\tilde{s}_B^{(1)}(L/2)& = 
C_F\left(L^2 - 2L +\frac{\pi^2}{6}+4\right) \nonumber \\
\tilde{s}_B^{(2)}(L/2)& = 
\frac89 L^4+\left(-\frac{76}{9}+ \frac{8}{27}n_l\right) L^3+
\left(\frac{496}{9}-\frac{28\pi^2}{27}-\frac{64}{27}n_l\right)L^2 \nonumber \\
&+\left(-\frac{4760}{27}+\frac{56\pi^2}{27}+40\zeta_3 
+  \frac{752}{81}n_l\right)L
+\frac{24824}{81}+\frac{19\pi^2}{3}\nonumber \\
& -\frac{143\pi^4}{405}-\frac{404}{9}\zeta_3 +
\left(-\frac{4496}{243}-\frac{10\pi^2}{81}+\frac{8}{27}\zeta_3\right)n_l \,.
\end{align}

We next collect expansion coefficients for 
the anomalous dimensions appearing in the RG equations in Section~\ref{sec:resum}. We define these as 
\begin{align}
  \label{eq:expsmallg}
  \gamma_{\cusp}(\alpha_s) = \frac{\alpha_s}{4 \pi} \left[
    \gamma^{\cusp}_0 + \left(\frac{\alpha_s}{4 \pi} \right)
    \gamma^{\cusp}_1 + \left(\frac{\alpha_s}{4 \pi} \right)^2
    \gamma^{\cusp}_2 + {\mathcal O}(\alpha_s^3) \right] \, .
\end{align}
and similarly for the other anomalous dimensions.  
One has \cite{Moch:2004pa}
\begin{align}
\gamma_0^{\rm cusp} &= 4 \,, \nonumber
  \\
  \gamma_1^{\rm cusp} &= \left( \frac{268}{9} - \frac{4\pi^2}{3}
  \right) C_A - \frac{80}{9}\,T_F n_l \,, \nonumber
  \\
  \gamma_2^{\rm cusp} &= C_A^2 \left( \frac{490}{3} -
    \frac{536\pi^2}{27} + \frac{44\pi^4}{45} + \frac{88}{3}\,\zeta_3
  \right) + C_A T_F n_l \left( - \frac{1672}{27} + \frac{160\pi^2}{27}
    - \frac{224}{3}\,\zeta_3 \right) \nonumber
  \\
  & \mbox{}+ C_F T_F n_l \left( - \frac{220}{3} + 64\zeta_3 \right) -
  \frac{64}{27}\,T_F^2 n_l^2 \,. \label{eq:cusp}
\end{align}
For $\gamma^S$ (\ref{eq:RGEsD}) one finds
\cite{Gardi:2005yi,Becher:2005pd,Neubert:2007je}
\begin{align}
  \gamma^S_0 &= -2 C_F \, ,\nonumber
  \\
  \gamma^S_1 &= C_F \left[ \left( \frac{110}{27} + \frac{\pi^2}{18} -
      18 \zeta_3 \right) C_A + \left( \frac{8}{27} + \frac{2}{9}
      \pi^2\right) T_F n_l \right] \, .
\end{align}
Similarly, the coefficients of the expansion of $\gamma^B$ in (\ref{eq:RGEsB}) are \cite{Jain:2008gb}
\begin{align}
\gamma^B_0 &= 2 C_F \, ,\nonumber \\
\gamma^B_1 &=   C_F C_A \left(\frac{698}{27} - \frac{23}{18} \pi^2 - 10 \zeta_3\right) + 
   C_F T_f n_l \left(-\frac{232}{27} + \frac{2}{9} \pi^2\right)\, .
\end{align}
The PDF anomalous dimensions are 
\begin{align}
  \gamma^{\phi_q}_0 &= 3 C_F \, , \nonumber
  \\
  \gamma^{\phi_q}_1 &= C_F^2 \left( \frac{3}{2} \!-\! 2 \pi^2 \!+\!24
    \zeta_3 \right) \!+\! C_F C_A \left(\frac{17}{6} \!+\! \frac{22
      \pi^2}{9} \!-\! 12 \zeta_3 \right) \!-\! C_F T_F n_l
  \left(\frac{2}{3} \!+\! \frac{8 \pi^2}{9} \right) \, ,
\end{align}
and 
\begin{align}
  \gamma^{\phi_g}_0 &= \frac{11}{3}C_A - \frac{4}{3}T_F n_l \, ,
  \nonumber
  \\
  \gamma^{\phi_g}_1 &= C_A^2 \left( \frac{32}{3} + 12 \zeta_3 \right)
  - \frac{16}{3} C_A T_F n_l - 4 C_F T_F n_l \, .
\end{align}
for the gluon and quark PDFs respectively.

In the $s$-channel singlet-octet basis, the matrix $\bm{\gamma}^h (x_t,\alpha_s)$ appearing in  the definition of $\bm{\gamma}^s (x_t,\alpha_s)$ (\ref{eq:gs}) is
 \begin{align}
 \bm{\gamma}^h_{q \bar{q}} (x_t,\alpha_s) &= 4 \gamma^q (\alpha_s)\bm{1} + N_c \gamma_{\cusp}(\alpha_s) \left( \ln x_t + i \pi\right)
 \left( \begin{array}{cc}
 0 & 0 \\ 
 0 & 1
 \end{array} \right)+ 2 \gamma_{\cusp}(\alpha_s)  \ln \frac{x_t}{1-x_t} \left(\begin{array}{cc}
0  & \frac{C_F}{2 N_c} \\ 
 1 & -\frac{1}{N_c} 
 \end{array}  \right) \, , \label{eq:ghqq}
 \end{align}
in the quark annihilation channel, while in the gluon fusion channel one finds
\begin{align}
\bm{\gamma}^h_{g g} (x_t,\alpha_s) &= 2\left( \gamma^g(\alpha_s) + \gamma^q (\alpha_s)\right)\bm{1} +N_c \gamma_{\cusp}(\alpha_s) \left( \ln x_t + i \pi\right)
 \left( \begin{array}{ccc}
 0 & 0 & 0\\ 
 0 & 1 & 0\\
 0 & 0 & 1
 \end{array} \right) \nonumber \\
& + 2 \gamma_{\cusp}(\alpha_s)  \ln \frac{x_t}{1-x_t} \left(\begin{array}{ccc}
0  & \frac{1}{2} & 0  \\ 
 1 & -\frac{N_c}{4} & \frac{N_c^2-4}{4 N_c} \\
0  & \frac{N_c}{4} & -\frac{N_c}{4}
 \end{array}  \right) \, . \label{eq:ghgg}
\end{align}
The anomalous dimensions $\gamma^q$ and $\gamma^g$, entering in  (\ref{eq:ghqq}, \ref{eq:ghgg}) and, consequently  in $\bm{\gamma}^s$ in (\ref{eq:gs}), are \cite{Moch:2005id, Becher:2006mr}
\begin{align}
  \gamma_0^q &= -3 C_F \,, \nonumber
  \\
  \gamma_1^q &= C_F^2 \left( -\frac{3}{2} + 2\pi^2 - 24\zeta_3 \right)
  + C_F C_A \left( - \frac{961}{54} - \frac{11\pi^2}{6} + 26\zeta_3
  \right) + C_F T_F n_l \left( \frac{130}{27} + \frac{2\pi^2}{3}
  \right) ,
\end{align}
and \cite{Moch:2005id, Becher:2007ty}
\begin{align}
  \gamma_0^g &= - \frac{11}{3}\,C_A + \frac{4}{3}\,T_F n_l \,,
  \nonumber
  \\
  \gamma_1^g &= C_A^2 \left( -\frac{692}{27} + \frac{11\pi^2}{18} +
    2\zeta_3 \right) + C_A T_F n_l \left( \frac{256}{27} -
    \frac{2\pi^2}{9} \right) + 4 C_F T_F n_l \,.
\end{align}

Finally, we define expansion coefficients for the QCD $\beta$ function
as
\begin{align}
  \beta(\alpha_s) &= -2\alpha_s \left[ \beta_0 \,
    \frac{\alpha_s}{4\pi} + \dots \right]
    \end{align}
where we need only
\begin{align}
  \beta_0 &= \frac{11}{3} C_A - \frac{4}{3} T_F n_l \,.
  \end{align}

\end{document}